\documentclass[10pt,3p,times]{elsarticle}

\usepackage{amsmath,amsfonts,amssymb}
\usepackage{amsthm}
\usepackage{bm}
\usepackage{booktabs}
\usepackage[toc]{appendix}
\usepackage[T1]{fontenc}
\usepackage{textcomp}
\usepackage{graphicx,epstopdf}
\usepackage[position=top,labelfont=normalfont,textfont=normalfont,singlelinecheck=off,justification=raggedright]{subfig}
\usepackage{floatrow}
\usepackage[dvipsnames]{xcolor}
\usepackage{tcolorbox}
\usepackage{setspace}
\usepackage{enumerate,etaremune}
\usepackage{tabularx,multirow}
\usepackage{framed}
\usepackage{hhline}
\usepackage[unicode,bookmarks=false]{hyperref}
\usepackage{url}
\hypersetup{
  colorlinks,
  citecolor=Blue,linkcolor=Blue,urlcolor=Blue
}
\usepackage{lineno}

\usepackage{tikz}
\usetikzlibrary{shapes.geometric}
\usetikzlibrary{shapes,arrows}
\usetikzlibrary{plotmarks}

\usetikzlibrary{calc}

\usepackage{algorithm}
\usepackage{algorithmicx,algpseudocode}
\usepackage{cases}

\usepackage[textsize=footnotesize,linecolor=red,backgroundcolor=red!25,bordercolor=red]
  {todonotes}

\newcommand{\eg}{{\it e.g.}}
\newcommand{\ie}{{\it i.e.}}

\newcommand{\etal}{{\it et al.}}

\newcommand{\stress}{\sigma}
\newcommand{\strain}{\varepsilon}
\newcommand{\tstress}{\bm{\stress}}
\newcommand{\tstrain}{\bm{\strain}}

\newcommand{\pd}{\partial}
\newcommand{\rDelta}{\mathrm{\Delta}}
\newcommand{\el}{\mathrm{e}}

\newcommand{\vpl}{\mathrm{vp}}
\newcommand{\trial}{\mathrm{tr}}
\newcommand{\romannumber}[1]{\uppercase\expandafter{\romannumeral #1\relax}}
\newcommand{\soft}{\mathrm{s}}
\newcommand{\hard}{\mathrm{h}}

\DeclareMathOperator{\Diver}{div}

\DeclareMathOperator{\sym}{sym}

\DeclareMathOperator{\tr}{tr}

\DeclareMathOperator{\dyad}{\otimes}
\newsavebox{\dotbox}

\theoremstyle{remark}
\newtheorem{remark}{Remark}

\newcommand{\revised}[1]{{\color{black} #1}}

\newcolumntype{L}[1]{>{\raggedright\let\newline\\arraybackslash\hspace{0pt}}m{#1}}
\newcolumntype{C}[1]{>{\centering\let\newline\\arraybackslash\hspace{0pt}}m{#1}}
\newcolumntype{R}[1]{>{\raggedleft\let\newline\\arraybackslash\hspace{0pt}}m{#1}}

\allowdisplaybreaks

\biboptions{sort&compress,square,comma,numbers}
\AtBeginDocument{\hypersetup{citecolor=RoyalBlue,linkcolor=RoyalBlue,urlcolor=RoyalBlue}}

\usepackage{etoolbox}
\makeatletter
\patchcmd{\ps@pprintTitle}
{Preprint submitted to}
{~}
{}{}
\makeatother

\begin{document}

\begin{frontmatter}

\title{An anisotropic viscoplasticity model for shale based on\\ layered microstructure homogenization}

\author[HKU]{Jinhyun Choo\corref{corr}} \ead{jchoo@hku.hk}
\author[UCSD]{Shabnam J. Semnani}
\author[LLNL]{Joshua A. White}

\cortext[corr]{Corresponding Author}

\address[HKU]{Department of Civil Engineering, The University of Hong Kong, Hong Kong}
\address[UCSD]{Department of Structural Engineering, University of California, San Diego, United States}
\address[LLNL]{Atmospheric, Earth, and Energy Division, Lawrence Livermore National Laboratory, United States}

\journal{~}

\begin{abstract}
Viscoplastic deformation of shale is frequently observed in many subsurface applications.
Many studies have suggested that this viscoplastic behavior is anisotropic\revised{---specifically, transversely isotropic---}and closely linked to the layered composite structure at the microscale.
In this work, we develop a two-scale constitutive model for shale in which anisotropic viscoplastic behavior naturally emerges from semi-analytical homogenization of a bi-layer microstructure.
The microstructure is modeled as a composite of soft layers, representing a ductile matrix formed by clay and organics, and hard layers, corresponding to a brittle matrix composed of stiff minerals.
This layered microstructure renders the macroscopic behavior anisotropic, even when the individual layers are modeled with isotropic constitutive laws.
Using a common correlation between clay and organic content and magnitude of creep, we apply a viscoplastic Modified Cam-Clay plasticity model to the soft layers, while treating the hard layers as a linear elastic material to minimize the number of calibration parameters.
We then describe the implementation of the proposed model in a standard material update subroutine.
The model is validated with laboratory creep data on samples from three gas shale formations.
We also demonstrate the computational behavior of the proposed model through simulation of time-dependent borehole closure in a shale formation with different bedding plane directions.
\end{abstract}


\end{frontmatter}

\section{Introduction}
\label{sec:intro}

Shale often accumulates a significant amount of irrecoverable deformation over time.
This viscoplastic deformation plays an important role in a variety of subsurface practices, such as in-situ stress estimation~\cite{Sone2014a,Sone2014b}, borehole drilling~\cite{Horsrud1994}, leakage prevention~\cite{Cerasi2017,Xie2019}, and hydraulic fracturing~\cite{Asala2016,Yang2016,Chau2017}.
This importance has motivated a number of experimental investigations into creep of shale at multiple scales (\eg~\cite{Chang2009,Li2012CreepShale,Sone2013b,Bennett2015,Rybacki2017,geng2018time,rassouli2018comparison,Trzeciak2018,Slim2019,Mighani2019CreepExperiments,Herrmann2020,Li2020}).

\begin{figure}[htbp]
  \centering
  \includegraphics[width=\textwidth]{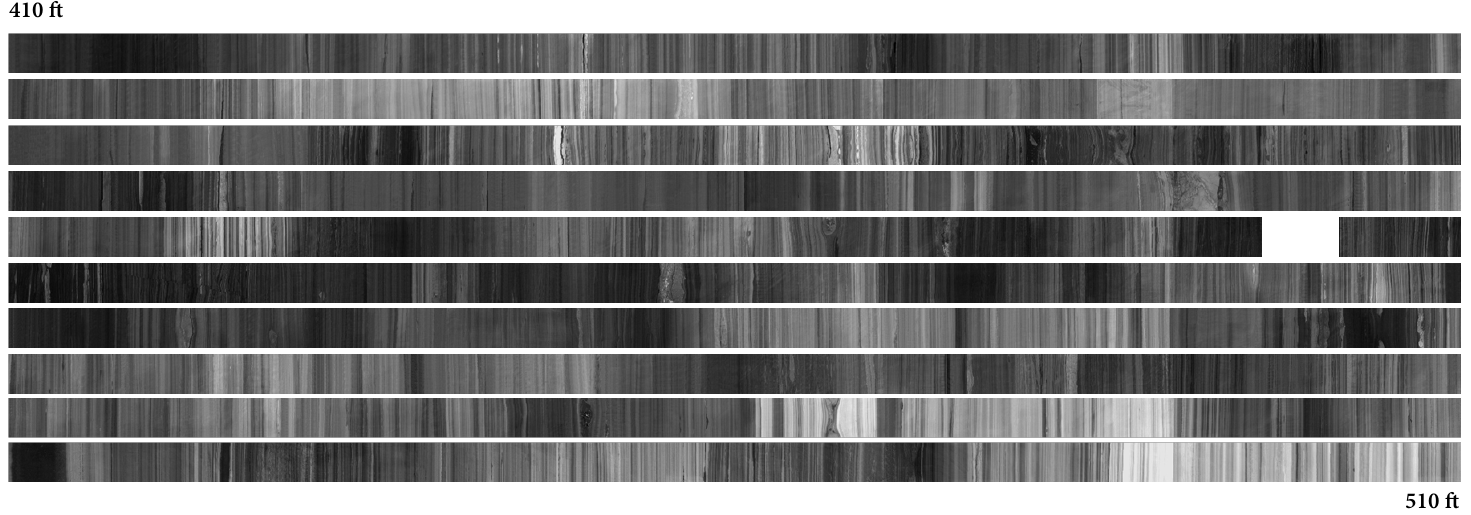}
  \caption{Optical log of core from the Green River Formation \cite{Mehmani2016}. The log samples a 100 ft (30.5 m) vertical interval through the Mahogany Zone, illustrating the strongly bedded nature of shales. Darker sections contain higher volume fractions of organic content than lighter sections. (Figure adapted from Mehmani \etal~\cite{Mehmani2016} with permission from Elsevier.) }
  \label{fig:layered-shale}
\end{figure}

Experimental results have shown that the viscoplastic behavior of shale is anisotropic---specifically, transversely isotropic---and closely linked to the shale composition.
For example, creep responses of samples from various shale formations have commonly exhibited a strong dependence on the relative direction between the differential stress and the bedding plane (\eg~\cite{Sone2013b,Sone2014a,Sone2014b,Li2020}).
\revised{Such anisotropy is a natural consequence of the fact that layering is ubiquitous in sedimentary rocks, including shale---see, \eg,~Fig.~\ref{fig:layered-shale}.}
Sone and Zoback~\cite{Sone2013b} have also found that creep behavior is apparently correlated with the elastic moduli of shale, although the mechanisms underlying viscoplastic and elastic deformations are quite different.
They explain this correlation by appealing to stress partitioning in a conceptual bi-layer model, in which one layer is composed of soft constituents such as clay and organics (kerogen), while the other layer is composed of hard constituents such as quartz, feldspar, pyrite (QFP), and carbonates.
This bi-layer representation is a simple but promising approach.
\revised{It is well justified by the fact that shale creep is mainly attributed to the clay and organic constituents rather than the hard minerals~\cite{Chang2009,Sone2013b,Rybacki2017,Trzeciak2018,Slim2019,Herrmann2020}, and it naturally captures the transverse isotropy resulting from the depositional process.}

Mathematically, time-dependent deformation of geomaterials has been described by viscoelastic/plastic continuum models (\eg~\cite{Trzeciak2018Long-termPoland,Sone2014b}) or empirical relations (\eg~\cite{DUSSEAULT1993Time-dependentRocks}).
For clay-rich materials, a number of anisotropic viscoplastic constitutive models (\eg~\cite{Zhou2005ElasticTests,Sivasithamparam2015ModellingSoils,Jiang2017EvaluationClays,Leoni2008AnisotropicSoils,Yin2010AnClays,Karim2014ReviewClays}) have been proposed by extending the Modified Cam-Clay (MCC) model originally developed for isotropic materials \cite{roscoe1968}.
Although most of these models have treated anisotropy in a purely macroscopic manner, without explicit consideration of material microstructure, a few models have incorporated anisotropy based on microstructural approaches.
For example, Pietruszczak \etal~\cite{Pietruszczak2004DescriptionMaterials} have proposed a constitutive framework that introduces two microstructure-based quantities, namely, a microstructure evolution parameter capturing creep deformation associated with the rearrangement of microstructure, and a microstructure tensor describing the direction of transverse isotropy.
Very recently, Borja \etal~\cite{Borja2020} have developed a two-material framework that represents shale as a mixture of a softer frame and a harder frame, applying an anisotropic critical state plasticity model~\cite{Semnani2016} to the harder frame to simulate anisotropic viscoplastic deformation.

Outside the geomechanics community, homogenization techniques have been developed as an elegant way to incorporate microstructure effects on viscoelastic \cite{Yu2002MultiscaleProblem,Lahellec2007EffectiveApproach} and viscoplastic \cite{Fotiu1996OverallComposites,Nesenenko2007HomogenizationViscoplasticity,Nesenenko2013HomogenizationType} behavior of materials.
However, the application of homogenization techniques to viscoplastic materials entails a few major difficulties.
First, the mathematical description of the macroscopic behavior cannot be determined a priori, because the macroscopic viscous behavior does not necessarily follow the same law as the constituents \cite{Mercier2009HomogenizationSchemes}. For example, a Maxwell-type behavior for the microscopic constituents does not necessarily lead to an overall macroscopic behavior following the Maxwell law \cite{Suquet1987ElementsMechanics}. Second, exact interactions between constituents can only be derived for linear elastic behavior, while new assumptions and strategies are needed for viscoplastic models.

A survey of mathematical homogenization techniques for dissipative materials can be found in Charalambakis \etal~\cite{Charalambakis2018MathematicalProgress}.
The majority of homogenization techniques developed for viscoplastic creep are based on mean-field methods \cite{Berbenni2015AExtension}, self-consistent schemes \cite{Paquin1999IntegralMaterials}, and transformation field analysis \cite{Dvorak1992TransformationMaterials}. These methods were initially developed for linear elastic composites based on Eshelby's solution for an ellipsoidal inclusion in an infinite matrix \cite{Eshelby1961}, and were later extended to the nonlinear regime by linearization of the local constitutive equations and definition of a linear comparison composite for a given deformation state \cite{Abou-ChakraGuery2008AGeomaterial,Pierard2006AnComposites,Doghri2010Mean-fieldMethod,Lahellec2007EffectiveApproach,Lahellec2007OnPrinciples,Brassart2012HomogenizationPrinciple,Boudet2016AnComposites,Brassart2011HomogenizationApproaches}. In particular, interaction laws for elasto-viscoplastic inclusions were developed to extend the Mori--Tanaka and self-consistent schemes \cite{Molinari2002AveragingMaterials,Mercier2009HomogenizationSchemesb,Mareau2009MicromechanicalMaterials,Paquin1999IntegralMaterials,Mareau2015AnMethod,Sabar2002AMaterials}. Specific assumptions such as piecewise uniform \cite{Marfia2018MultiscaleComposites} or non-uniform \cite{Covezzi2017HomogenizationTFA} distribution of the strain field in each phase were used to accommodate viscoplasticity within the transformation field analysis framework \cite{Roussette2009NonuniformComposites,Kruch2011Multi-scaleDamage,Barral2020HomogenizationValidation}.

Other classes of homogenization methods include asymptotic expansion and periodic homogenization methods. Periodic homogenization expresses the strain field as the sum of a macroscopic field and a periodic perturbation \cite{Suquet1987ElementsMechanics}. This technique was applied to elasto-viscoplastic composites under simplifying assumptions such as uniform macroscopic stress and strain in the unit cell \cite{Ohno2000HomogenizedStructures,Nakata2008Multi-scaleConfigurationb} and point symmetry of internal distributions \cite{Ohno2001ADistributions}. Asymptotic expansion, which forms the basis of the present work, has been applied to viscoelasticity \cite{Yu2002MultiscaleProblem,Chung2000AMedia} and viscoplasticty \cite{Fish1998ComputationalHomogenization} by assuming viscoplastic strains are piecewise constant eigenstrains, \ie~strains not caused by external stresses.

The above techniques are generally applicable to isotropic composites made of ellipsoidal inclusions within a matrix, while homogenization-based anisotropic viscoplasticity has received little attention in the literature \cite{Chatzigeorgiou2016PeriodicMaterialsb,Matsuda2002HomogenizedLaminates,Shamaev2016HomogenizationMaterials,Semnani2020}. Chatzigeorgiou \etal~\cite{Chatzigeorgiou2016PeriodicMaterialsb} applied asymptotic expansion techniques to composites with generalized standard material laws and demonstrated their application to multi-layered media. Mathematical approaches have also been adopted for perfectly bonded isotropic creep materials \cite{Shamaev2016HomogenizationMaterials} and fiber-reinforced laminates  \cite{Matsuda2002HomogenizedLaminates}.

In this work, we develop a two-scale constitutive model for shale in which anisotropic viscoplastic behavior naturally emerges from semi-analytical homogenization of a bi-layer microstructure.
The microstructure is a composite of soft layers, representing a ductile matrix formed by clay and organics, and hard layers, corresponding to a brittle matrix composed of QFP and carbonates, similar to the conceptual model of Sone and Zoback~\cite{Sone2013b}.
In particular, we build the model using an inelastic homogenization framework for layered materials recently proposed by Semnani and White~\cite{Semnani2020}, here extending it to viscoplastic materials.
One of the main advantages of this homogenization approach is that the macroscopic behavior is inherently anisotropic, even when the individual layers are modeled with isotropic constitutive laws.  This allows
us to adopt standard isotropic models with physically meaningful parameters.
This provides a significant advantage over many single-scale models that require anisotropy parameters that can be challenging to calibrate and physically justify.
Using a common correlation between clay and organic content and magnitude of creep, we apply a viscoplastic Modified Cam-Clay plasticity model to the soft layers, while treating the hard layers as a linear elastic material.
It will be shown that the proposed model can simulate viscoplastic deformation of shale remarkably well with a relatively small number of parameters.
Through a borehole closure example, we will also demonstrate that the model can be efficiently applied to general initial--boundary value problems involving anisotropic deformation of shale over time.

The proposed model is distinguished from existing shale models in several aspects.
First, the majority of existing viscoplasticity models for shale assume isotropic behavior \cite{Haghighat2020,Xu2017UtilizingBasin}.
Although an isotropic viscoplasticity model may reproduce uniaxial creep deformation, it requires a different parameterization when the bedding plane direction is changed; see Haghighat \etal~\cite{Haghighat2020} for example.
Second, it differs from the anisotropic viscoplasticity model of shale recently proposed by Borja \etal~\cite{Borja2020}. That work relies on solid--solid mixture theory, while the present work employs asymptotic expansion.
Both represent shale as a composite of harder and softer materials.  In Borja \etal~\cite{Borja2020}, however, the materials are not layered, requiring at least one of them be modeled by a single-scale anisotropic constitutive law.
The model proposed here simplifies calibration, since anisotropic properties naturally emerge from the layered microstructure.
Third, the proposed model differs from other homogenization-based models in that it provides a general and flexible framework to accommodate any desired behavior for the constituents (\eg~brittle vs. ductile, rate-dependent vs. rate-independent).  Its algorithmic implementation in a computational mechanics code is also quite straightforward.
Finally, the viscoplastic deformation considered in this work is an intrinsic process of the solid matrix, rather than induced by pore fluid diffusion or other physio-chemical processes studied in other works (\eg~\cite{navarro2001secondary,cosenza2014secondary,choo2016hydromechanical,borja2016cam}).

The remainder of the paper is organized as follows.
Section~\ref{sec:formulation} describes the model formulation.
Section~\ref{sec:implementation} presents algorithms for implementing the proposed model in a material update subroutine.
Section~\ref{sec:validation} validates the model with laboratory creep data of samples from three gas shale formations, namely, Haynesville, Eagle Ford, and Barnett.
Section~\ref{sec:example} demonstrates the computational performance of the model through numerical simulation of time-dependent borehole closure in a shale formation with different bedding plane directions.
Section~\ref{sec:closure} closes the work.

As for sign conventions and notations, stresses and strains are positive in compression following the geomechanics sign convention.
Bold-face letters denote tensors and vectors.
The symbol ``$\cdot$'' denotes single contraction, and ``$:$'' denotes double contraction.

\section{Formulation}
\label{sec:formulation}

This section begins by describing a bi-layer model of shale, which is a particular class of the inelastic homogenization framework developed by Semnani and White~\cite{Semnani2020}.
For brevity, the complete derivation of the homogenization framework will be omitted, as it is extensively described in the prior work.
Specific constitutive models will then be introduced to model the stress-strain response of the two layers at the microscale, aimed at capturing viscoplastic deformation in shale with minimal ingredients.
Quasi-static behavior and infinitesimal deformation will be assumed throughout.

\subsection{Bi-layer model}
\begin{figure}[htbp]
    \centering
    \includegraphics[width=0.55\textwidth]{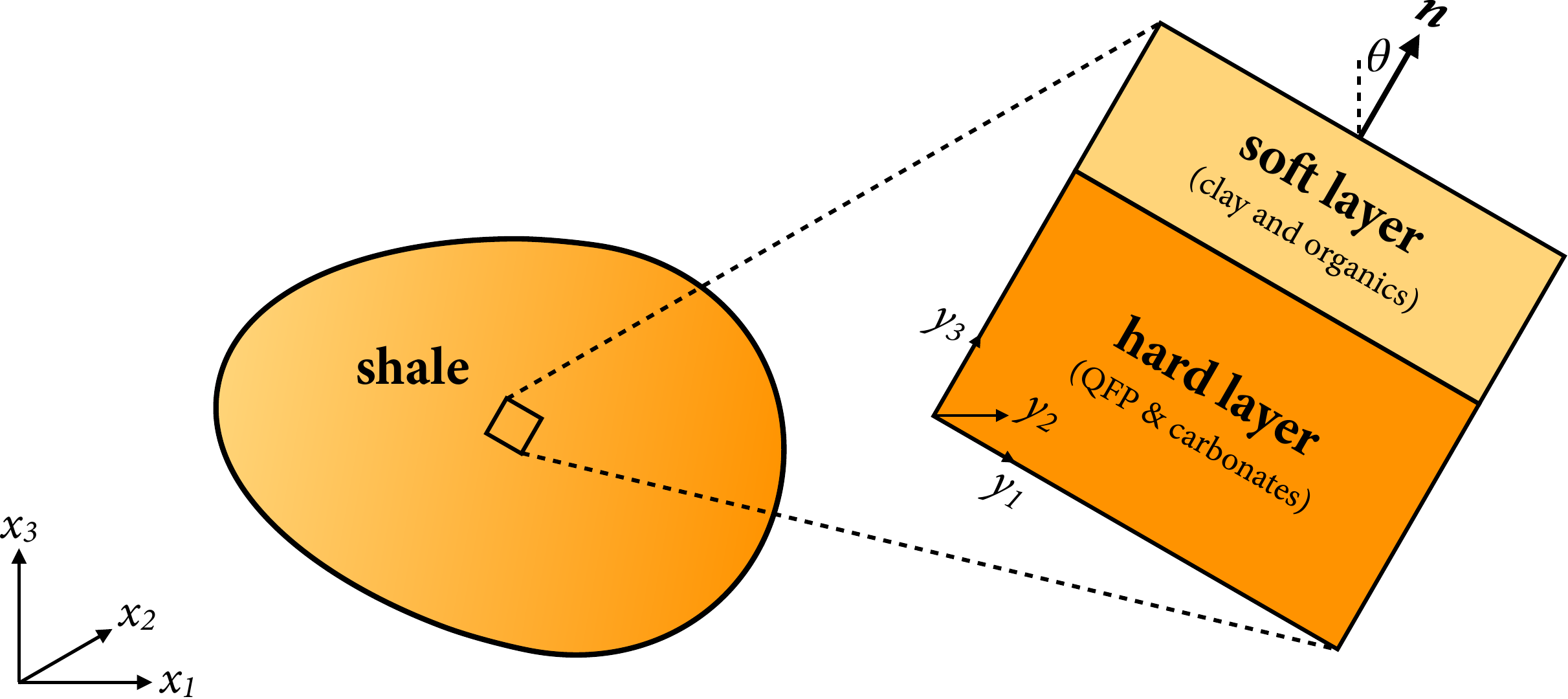}
    \caption{An illustration of a unit cell comprised of a soft layer and a hard layer. The macroscopic coordinate system is $\bm{x}=\{x_1, x_2, x_3\}$, and the microscopic coordinate system is $\bm{y}=\{y_1,y_2, y_3\}$. Symbols $\bm{n}$ and $\theta$ denote the unit normal vector of the bedding plane and its angle from the vertical, respectively.}
    \label{fig:bi-layer-unitcell}
\end{figure}

Consider a body $\mathcal{B}$ whose microstructure is described by a spatially periodic unit cell $\mathcal{U}$, as illustrated in Fig.~\ref{fig:bi-layer-unitcell}.
The unit cell is composed of two parallel layers, namely, a soft layer and a hard layer.
The soft layer represents the compliant constituents of shale, such as clay and organic matters.
The hard layer corresponds to the stiff constituents of shale, including quartz, feldspar, pyrite (QFP) and carbonate.

A few assumptions are introduced to this bi-layer model.
First, the characteristic length-scale of the body (macroscale) and the unit cell (microscale) are clearly separated, a key requirement under asymptotic homogenization theory.
Second, each layer is a homogeneous, standard continuum.
Third, the two layers are perfectly bonded to each other, such that the displacement field is continuous across the layers as well as within the layers.
Note, however, that certain strain components may be discontinuous at the layer interface.

From a practical point of view, we also note that only \emph{quasi-periodicity} of the microstructure is strictly required.  The material properties of the medium may vary slowly over large-length scales, as long as the local properties at short scales may be well-approximated via a periodic unit cell.  In a boundary value problem, for example, the microstructure at each integration point is assumed to be periodic, but the material properties assigned from one integration point to the next may vary slowly to capture large-scale heterogeneities like those seen in Fig.~\ref{fig:layered-shale}.

To describe the kinematics of the two-scale model, we introduce a macroscopic coordinate system, $\bm{x}$, and a microscopic coordinate system, $\bm{y}$.
The two coordinate systems are related as
\begin{align}
    \bm{y} = \bm{x}/\epsilon\,,
\end{align}
where $\epsilon \ll 1$ is the period of the unit cell.
We also define a scalar micro-coordinate axis $y:=\bm{y}\cdot\bm{n}$, where $\bm{n}$ is the unit normal vector of the layers.
Let us denote by $\bm{u}(\bm{x},y)$ the displacement field, with the dual argument emphasizing that this field is a function of both the macroscopic and microscopic coordinates.
Applying two-scale asymptotic homogenization,
we approximate $\bm{u}(\bm{x},y)$ as a truncated power series of $\epsilon$, \ie~
\begin{align}
    \bm{u}(\bm{x},y) \approx \bm{u}^{(0)}(\bm{x}) + \epsilon\,\bm{u}^{(1)}(\bm{x},y)\,.
\end{align}
Here, $\bm{u}^{(0)}$ represents the macroscopic component of the displacement field, which is only a function of the macroscopic coordinate.  In contrast, $\bm{u}^{(1)}$ represents a microscopic displacement fluctuation, which varies quasi-periodically across unit cells.
Accordingly, the total strain field can be additively decomposed as
\begin{align}
    \tstrain = \bm{E} + \bm{e}\,,
\end{align}
where $\bm{E}$ and $\bm{e}$ are the macroscopic and microscopic parts of the strain field, given by
\begin{align}
    \bm{E} := \sym\left(\frac{\pd\bm{u}^{(0)}}{\pd\bm{x}}\right ) \,, \quad
    \bm{e} := \sym\left(\frac{\pd\bm{u}^{(1)}}{\pd y}\dyad\bm{n}\right)\,,
    \label{eq:strains}
\end{align}
with $\sym(\cdot)$ denoting the symmetric part of a tensor.
The strain field is potentially discontinuous at the layer interfaces, despite the continuity of displacement field throughout the unit cell. In Semnani and White~\cite{Semnani2020} the possibility of inter-layer slip or normal opening is considered, but here we focus on perfectly bonded layers.
The strain field must then be continuous in the direction tangential to the layers.

The governing equations of the boundary value problem are derived as in Semnani and White~\cite{Semnani2020}.
This procedure will be omitted for brevity.
To express the equations, let us introduce layer-wise quantities, using subscripts $(\cdot)_{\soft}$ and $(\cdot)_{\hard}$ to denote quantities in the soft and hard layers, respectively.
We denote by $\phi_{\soft}$ and $\phi_{\hard}$ the volume fractions of the soft and hard layers, which satisfy $\phi_{\soft}+\phi_{\hard}=1$.
We also define the microscopic displacement gradient vectors of the two layers as
\begin{align}
    \bm{v}_{\soft} := \frac{\pd\bm{u}^{(1)}_{\soft}}{\pd y}\,, \quad
    \bm{v}_{\hard} := \frac{\pd\bm{u}^{(1)}_{\hard}}{\pd y}\,.
\end{align}
To satisfy the bedding-normal traction continuity requirement, we note that the microscopic displacement gradients must be constant in each layer \cite{Semnani2020}.  The total strain tensors of the layers are then defined as
\begin{align}
    \tstrain_{\soft} := \bm{E} + \sym(\bm{v}_{\soft}\dyad\bm{n})\,, \quad
    \tstrain_{\hard} := \bm{E} + \sym(\bm{v}_{\hard}\dyad\bm{n})\,.
    \label{eq:micro-strains}
\end{align}
These strains are related to the stresses in the individual layers as
\begin{align}
    \dot\tstress_{\soft} = {\mathbb{C}}_{\soft}:\dot{\tstrain}_{\soft}\,, \quad
    \dot\tstress_{\hard} = {\mathbb{C}}_{\hard}:\dot{\tstrain}_{\hard}\,,
    \label{eq:micro-constitutive-law}
\end{align}
where $\tstress$ and $\mathbb{C}$ denote the stress tensor and the stress--strain tangent tensor within each phase, respectively.
The general rate form is intended to accommodate potentially nonlinear material behavior.  The particular models chosen for the layers do not impact to the overall homogenization algorithm, so we leave them unspecified for the moment.  We denote by $\bm{\Sigma}$ the macroscopic stress tensor, which is defined as the volume average of the stress tensors in each layer, \ie
\begin{align}
    \bm{\Sigma} := \phi_{\soft}\tstress_{\soft} + \phi_{\hard}\tstress_{\hard}\,.
\end{align}

Introducing the multiscale expansions into the global initial-boundary-value problem, we may arrive at two separate but coupled problems: a \emph{microscale problem} in the unit cell $\mathcal{U}$ and a \emph{macroscale problem} in the body $\mathcal{B}$.  The governing equations of the microscale problem enforce certain stress and displacement field compatibilities:
\begin{align}
    \tstress_{\soft}\cdot\bm{n} - \bm{t} &= \bm{0} \quad \text{in}\; \mathcal{U}_s \quad \text{(traction continuity in the soft layer)}\,,
    \label{eq:micro-equilibrium-soft} \\
    \tstress_{\hard}\cdot\bm{n} - \bm{t} &= \bm{0} \quad \text{in}\; \mathcal{U}_h\quad \text{(traction continuity in the hard layer)}\,,
    \label{eq:micro-equilibrium-hard} \\
    \phi_{\soft}\bm{v}_{\soft} + \phi_{\hard}\bm{v}_{\hard} &= \bm{0} \quad \text{in}\; \mathcal{U} \; \quad \text{(periodicity of micro-displacements)}
    \label{eq:micro-compatibility} \,.
\end{align}
Here, $\bm{t}= \bm{\Sigma}\cdot\bm{n}$ denotes the traction vector resulting from stresses acting in the bedding normal direction, which must be continuous in all layers to satisfy unit cell equilibrium.

The governing equation of the macroscopic problem is identical to the linear momentum balance typically employed in single-scale models,
\begin{align}
    \Diver\,\bm{\Sigma} + \bm{F} = \bm{0} \quad \text{in}\; \mathcal{B}\,.
    \label{eq:macro-governing-eq}
\end{align}
Here, $\Diver(\cdot)$ is the divergence operator with respect to the macroscopic coordinate system $\bm{x}$,
and $\bm{F}$ is the body force vector.  This governing equation is supplemented with appropriate initial conditions in $\mathcal{B}$ and boundary conditions on $\partial \mathcal{B}$ to define a complete problem.

\subsection{Soft layer: Viscoplastic Modified Cam-Clay}
We may now begin to specialize the general model by defining isotropic constitutive models for the layers.  For the soft layer composed of clay and organic content, we use a viscoplastic model.  This layer is then responsible for the time-dependent ductile deformation of the overall shale.
The viscoplastic formulation begins by writing the stress--strain relationship as
\begin{align}
    \dot{\tstress}_{\soft} = \mathbb{C}^{\el}_{\soft} : (\dot{\tstrain}_{\soft} - \dot{\tstrain}^{\vpl}_{\soft})\,,
\end{align}
where $\mathbb{C}^{\el}_{\soft}$ is the elastic stress--strain tangent of the soft layer,
and $\tstrain^{\vpl}_{\soft}$ is the viscoplastic part of the strain in the soft layer.
For simplicity, we assume that the elastic part of the soft layer deformation is linear elastic.
The linear elastic tangent can be written as
\begin{align}
    \mathbb{C}^{\el}_{\soft} = K_{\soft}(\bm{1}\dyad\bm{1}) + 2G_{\soft}\left(\mathbb{I} - \frac{1}{3}\bm{1}\dyad\bm{1} \right)\,,
\end{align}
with $K_{\soft}$ and $G_{\soft}$ denoting the bulk and shear moduli of the soft layer, respectively.
Also, $\bm{1}$ and $\mathbb{I}$ denote the second-order identity tensor and the fourth-order symmetric identity tensor, respectively.
While the elastic behavior of clay-rich materials can be highly nonlinear and pressure-dependent~\cite{Choo2011,Choo2013}, the assumption of linear elasticity allows us to minimize the number of model parameters when viscoplastic deformation is of primary interest.  We note, however, that the standard Modified Cam-Clay model typically employs a pressure-dependent elastic response rather than a linear model.

For the viscoplastic part of deformation, we adopt the modeling approach proposed by Duvaut--Lions~\cite{Duvaut1972}, which is one of the most popular approaches to viscoplasticity in the literature (\eg~\cite{Simo1988,Ju1990,wang1997viscoplasticity,lazari2015local,Borja2020}).
\revised{The Duvaut--Lions viscoplasticity is chosen mainly for its relative simplicity and computational robustness.
As shown in Simo \etal~\cite{Simo1988}, the Duvaut--Lions approach can be easily applied to extend any rate-independent plasticity model to viscoplasticity, regardless of the smoothness of yield surface.
We note, however, that other well-established approaches to viscoplasticity, such as the Perzyna formulation~\cite{perzyna1966fundamental}, can also be cast into the present homogenization framework.
Indeed, Borja \etal~\cite{Borja2020} have shown that the Duvaut--Lions and Perzyna approaches can produce nearly the same viscoplastic behavior when their parameters are properly selected.
}

Central to the Duvaut--Lions viscoplasticity is the notion of backbone stress, which is the closest-point projection of stress onto the yield surface of a rate-independent plasticity model.
In a Duvaut--Lions model, the backbone stress tensor is first computed based on a rate-independent model,
followed by calculation of the (actual) stress tensor by integrating the rate equations accommodating viscous effects.

For the rate-independent model determining the backbone stress tensor, here we choose the Modified Cam-Clay model~\cite{roscoe1968}, which is widely used for clay-rich materials including shale (\eg~\cite{Chang2010,Semnani2016,white2017thermoplasticity,zhao2018strength}).
We denote by $p$ and $q$ the volumetric and deviatoric stress invariants, respectively, which are defined as
\begin{align}
    p := \frac{1}{3}\tr(\tstress_{\soft})\,,\quad
    q := \sqrt{\frac{2}{3}}\left\|\tstress_{\soft} - p\bm{1}\right\|.
\end{align}
Using these two stress invariants, the yield function can be written as
\begin{align}
    f(p,q,p_{c}) = \frac{q^{2}}{M^2} + p(p - p_{c}) \leq 0\,,
    \label{eq:yield}
\end{align}
where $M>0$ is the slope of the critical state line (CSL), and $p_{c}>0$ is the preconsolidation pressure which is the hardening variable of the model.
The hardening law is given by
\begin{align}
    \dot{p}_{c} = \frac{p_{c}}{\lambda}\dot{\strain}_{\rm v}^{\vpl}\,,
    \label{eq:mcc-hardening}
\end{align}
where $\strain_{\rm v}^{\vpl} := \tr(\tstrain^{\vpl})$ is the volumetric part of the viscoplastic strain.
As standard in the Modified Cam-Clay model, the plastic flow will be assumed to be associative.

The extension of this model to Duvaut--Lions viscoplasticity is carried out through the rate equations~\cite{Simo1988}
\begin{align}
    \dot{\tstrain}^{\vpl}_{\soft} &= \frac{1}{\tau}(\mathbb{C}^{\el}_{\soft})^{-1}:(\tstress_{\soft} - \bar{\tstress}_{\soft})\,, \label{eq:dv-stress-rate}\\
    \dot{\strain}_{\rm v}^{\vpl} &= -\frac{1}{\tau}\left(\frac{\lambda}{p_{c}}\right)(p_{c} - \bar{p}_{c})\,,
    \label{eq:dv-hardening-rate-v0}
\end{align}
where the bar denotes the inviscid part of a quantity calculated by closest-point projection onto the rate-independent yield surface.
\revised{Note that while Eq.~\eqref{eq:dv-stress-rate} is general for any plasticity model, Eq.~\eqref{eq:dv-hardening-rate-v0} is a particular case of the rate equation for internal variables~\cite{Simo1988} specialized to the Modified Cam-Clay model.
Substituting Eq.~\eqref{eq:mcc-hardening} into Eq.~\eqref{eq:dv-hardening-rate-v0} gives}
\begin{align}
    \dot{p}_{c} &= -\frac{1}{\tau}(p_{c} - \bar{p}_{c})\,.
    \label{eq:dv-hardening-rate}
\end{align}
Also, $\tau$ is a relaxation time parameter, which may be considered a constant or a function of other quantities as described in Simo \etal~\cite{Simo1988}.
In the context of isotropic Perzyna-type viscoplastic modeling of shale, Haghighat \etal~\cite{Haghighat2020} have proposed an exponential relationship between the viscosity coefficient and the volumetric viscoplastic strain $\strain_{\rm v}^{\vpl}$.
Adapting this relationship to Duvaut--Lions viscoplasticity, we consider the following form:
\begin{align}
    \tau = \tau_{0}\exp(\zeta \strain_{\rm v}^{\vpl})\,.
    \label{eq:relaxation-time}
\end{align}
Here, $\tau_{0}$ is the initial relaxation time, and $\zeta$ is a parameter determining the rate of evolution of the relaxation parameter.

\subsection{Hard layer: Linear elasticity}
The hard layer, which represents a matrix composed of QFP and carbonate minerals, typically deforms in a brittle and rate-independent manner.
Because brittle failure of the hard layer is outside the focus of this work, here we simply model the hard layer as a linear elastic material, reducing the overall number of material parameters.
Then the micro-stress tensor of the hard layer, $\tstress_{\hard}$, is related to the micro-strain tensor of the hard layer, $\tstrain_{\hard}$, as
\begin{align}
    \tstress_{\hard} = \mathbb{C}^{\el}_{\hard}:\tstrain_{\hard}\,,
\end{align}
where
\begin{align}
    \mathbb{C}^{\el}_{\hard} = K_{\hard}(\bm{1}\dyad\bm{1}) + 2G_{\hard}\left(\mathbb{I} - \frac{1}{3}\bm{1}\dyad\bm{1} \right)
\end{align}
is the elastic stress--strain tangent of the hard layer.
Again, $K_{\hard}$ and $G_{\hard}$ denote the bulk and shear moduli of the hard layer, respectively.

Note that if the brittle failure of the hard layer is also of interest, the linear elastic model should be replaced by a plasticity (or damage) model for brittle geomaterials.
This replacement can be made without any significant change to the homogenization procedure.
\smallskip

\begin{remark}
In the absence of viscoplastic deformation, the proposed model becomes equivalent to the bi-layer elastic model proposed by Backus~\cite{Backus1962}.
\end{remark}

\section{Implementation}
\label{sec:implementation}
This section describes how to implement the proposed two-scale model, with a focus on the implementation of the material update subroutine.  Its implementation in a continuum mechanics code is quite straightforward.
Consider a load step from time $t_{n}$ to $t_{n+1}$, in which the goal is to determine the macroscopic displacement field $\bm{u}_{n+1}$ for which both the macroscopic and microscopic problems are satisfied.  When using an implicit finite element method, we adopt a two level iteration: a global Newton's method is used to solve a discretized version of the macroscopic boundary value problem~\eqref{eq:macro-governing-eq}, while sub-iterations are used at each material point to solve the microscale balances~\eqref{eq:micro-equilibrium-soft}--\eqref{eq:micro-compatibility} to determine the local stress-strain response.

We begin by assuming that an estimate for the discrete macroscale displacement $\bm{u}^{(0)}_{n+1,\,m}$ is given, where subscript $(\cdot)_m$ is a global iteration counter.  The  macro-strain $\bm{E}$ at a quadrature point is therefore known via~\eqref{eq:strains} and used as an input to a material subroutine.  This routine uses the procedure prescribed below to determine the macroscopic stress $\bm{\Sigma}$ and tangent stiffness $\bm{C}_\Sigma$ satisfying the local equilibrium and compatibility equations.  While the internal computations differ, the strain-driven interface is the same as for any other material subroutine.  Global residual equations may then be assembled, and an updated displacement field $\bm{u}^{(0)}_{n+1,\,m+1}$ computed if necessary.  The procedure is iterated until all balance equations are satisfied to a desired tolerance.

In the following, we describe a procedure to numerically solve the microscale problem, which is a particular case of the solution algorithm presented in Semnani and White~\cite{Semnani2020}.  We briefly summarize the key steps here, but further details can be found in the prior work.
We also describe the integration procedure for the viscoplastic model of the soft layer.
For brevity, we will denote quantities at $t_{n+1}$ without an additional subscript, while distinguishing quantities at $t_{n}$ with a subscript $(\cdot)_{n}$.

\subsection{Solution to the microscale problem}
We also use Newton's method to solve the microscale problem, considering its potential nonlinearity due to viscoplasticity.
Newton's method for iteratively solving a system of nonlinear equations $\bm{R}(\bm{X})=\bm{0}$ proceeds in the two steps:
\begin{align}
    \text{solving} &\quad \bm{\mathcal{J}}^k \rDelta \bm{X} = - \bm{R}^k \,,\\
    \text{updating} &\quad \bm{X}^{k+1} = \bm{X}^{k} + \rDelta \bm{X} \,.
\end{align}
Here, $\bm{R}$ is the residual vector, $\bm{X}$ is the unknown vector, $\rDelta\bm{X}$ is the search direction, $\bm{\mathcal{J}}$ is the Jacobian matrix, and the superscript $(\cdot)^{k}$ is a local iteration counter.
For this particular problem, the residual vector consists of the three governing equations of the microscale problem, namely, Eqs.~\eqref{eq:micro-equilibrium-soft}--\eqref{eq:micro-compatibility},
and the unknown vector is an array of the three primary unknowns, namely, $\bm{v}_{\soft}$, $\bm{v}_{\hard}$, and $\bm{t}$.
Specifically,
\begin{align}
    \bm{\bm{R}}\left(\bm{X}\right)
    := \begin{bmatrix}
        \bm{n}\cdot\tstress_{\soft} -\bm{t}\\
        \bm{n}\cdot\tstress_{\hard} -\bm{t}\\
        \phi_{\soft} \bm{v}_{\soft} + \phi_{\hard} \bm{v}_{\hard}
    \end{bmatrix} \rightarrow \bm{0}\,, \quad
    \bm{X}
    := \begin{bmatrix}
        \bm{v}_{\soft} \\
        \bm{v}_{\hard} \\
        \bm{t}
    \end{bmatrix} \,.
    \label{eq:residual-unknown}
\end{align}
Note that the micro-stresses in the residual equations are a function of both the microscopic strains and the \emph{fixed} macroscopic strain via equations~\eqref{eq:micro-strains} and~\eqref{eq:micro-constitutive-law}.
The Jacobian matrix is given by
\begin{equation}
   \bm{\mathcal{J}}
   := \frac{\pd \bm{R}}{\pd \bm{X}}
   = \begin{bmatrix}
        \bm{n}\cdot\bm{\mathcal{C}}_{\soft}\cdot\bm{n} & & -\bm{1} \\
        & \bm{n}\cdot\bm{\mathcal{C}}_{\hard}\cdot\bm{n} & -\bm{1} \\
        \phi_{\soft} \bm{1} & \phi_{\hard} \bm{1} &
    \end{bmatrix} \,,
    \label{eq:Jacobian}
\end{equation}
where $\bm{\mathcal{C}}_{\soft}$ and $\bm{\mathcal{C}}_{\hard}$ are the consistent (algorithmic) tangent operators for the micro-constitutive laws in the soft and hard layers, respectively.
While $\bm{\mathcal{C}}_{\hard} = \mathbb{C}_{\hard}^{\el}$ for the linear elastic hard layer,
$\bm{\mathcal{C}}_{\soft}$ should be distinguished from the continuum (theoretical) tangent of the viscoplasticity model for an optimal convergence rate during the Newton iteration.
A specific expression for $\bm{\mathcal{C}}_{\soft}$ will be provided in Eq.~\eqref{eq:dv-cto-final} later in this section.

When the material update is performed within a global iteration algorithm (\eg~for a stress-driven problem or nonlinear finite element analysis),
the macroscopic tangent operator, $\pd\bm{\Sigma}/\pd\bm{E}$, is required for convergence of the global iteration.
The macroscopic tangent operator can be written as
\begin{align}
    \bm{\mathcal{C}}_{\Sigma}
    := \frac{\pd\bm{\Sigma}}{\pd\bm{E}}
    = \phi_{\soft}\frac{\pd\tstress_{\soft}}{\pd\bm{E}} + \phi_{\hard}\frac{\pd\tstress_{\hard}}{\pd\bm{E}}\,,
    \label{eq:macro-cto}
\end{align}
with
\begin{align}
    \frac{\pd\tstress_{\soft}}{\pd\bm{E}}
    &= \bm{\mathcal{C}}_{\soft} - (\bm{\mathcal{C}}_{\soft}\cdot\bm{n})\cdot\left(\bm{\mathcal{J}}^{-1}_{11}\cdot\bm{n}\cdot\bm{\mathcal{C}}_{\soft} + \bm{\mathcal{J}}^{-1}_{12}\cdot\bm{n}\cdot\bm{\mathcal{C}}_{\hard}\right), \label{eq:macro-cto-soft}\\
    \frac{\pd\tstress_{\hard}}{\pd\bm{E}}
    &= \bm{\mathcal{C}}_{\hard} - (\bm{\mathcal{C}}_{\hard}\cdot\bm{n})\cdot\left(\bm{\mathcal{J}}^{-1}_{21}\cdot\bm{n}\cdot\bm{\mathcal{C}}_{\soft} + \bm{\mathcal{J}}^{-1}_{22}\cdot\bm{n}\cdot\bm{\mathcal{C}}_{\hard}\right). \label{eq:macro-cto-hard}
\end{align}
Here, $\bm{\mathcal{J}}^{-1}_{(\cdot)(\cdot)}$ is the inverse of the Jacobian matrix~\eqref{eq:Jacobian} with subscripts denoting the appropriate sub-blocks of $\bm{\mathcal{J}}^{-1}$ (which has a 3$\times$3 block structure).
Notably, on the right hand sides of Eqs.~\eqref{eq:macro-cto-soft} and~\eqref{eq:macro-cto-hard},
the terms following the micro-tangents $\bm{\mathcal{C}}_{\soft}$ and $\bm{\mathcal{C}}_{\hard}$ have emerged from the interaction between the two layers.

A complete material update procedure is summarized in Algorithm~\ref{algo:material-update}.  Note that $\mathrm{tol}\ll 1$ is the tolerance for the Newton iteration.
Voigt representations of all quantities can be found in Semnani and White~\cite{Semnani2020}.
\smallskip

\begin{remark}
It should be noted that an additional level of nested iterations is required when using nonlinear material models for the layers (as here for the viscoplastic layer) because sub-iterations are required when calling the update routine for the layer stresses.   While deep nesting can be expensive, the material subroutine costs remain small compared to other components of an implicit finite element solution procedure, and they may be trivially parallelized.
\end{remark}

\begin{algorithm}[htbp]
    \setstretch{1.15}
    \caption{Material point update.}
    \begin{algorithmic}[1]
        \Require $\bm{E}$
        \Ensure $\bm{\Sigma}$ and $\bm{C}_{\Sigma}$
        \State Initialize the iteration counter $k=0$ and the unknown vector $\bm{X}^{k}=\bm{0}$.
        \State Calculate the total strains $\tstrain_{\soft} = \bm{E} + \sym(\bm{v}_{\soft}\dyad\bm{n})$ and $\tstrain_{\hard} = \bm{E} + \sym(\bm{v}_{\hard}\dyad\bm{n})$.
        \State Get $\tstress_{\soft}$, $\tstress_{\hard}$, $\bm{\mathcal{C}}_{\soft}$ and $\bm{\mathcal{C}}_{\hard}$ by passing $\tstrain_{\soft}$ and $\tstrain_{\hard}$ into material subroutines for the microscale constitutive models.
        \State Assemble the residual vector $\bm{R}^{k}(\bm{X}^{k})$ as Eq.~\eqref{eq:residual-unknown}.
        \If {$\|\bm{R}^{k}\| > \mathrm{tol}$}
            \State Solve for the search direction $\rDelta \bm{X} = -(\bm{\mathcal{J}}^k)^{-1}\bm{R}^k$.
            \State Perform a Newton update $\bm{X}^{k+1} = \bm{X}^{k} + \rDelta \bm{X}$.
            \State Set $k\leftarrow k+1$ and return to Step 2.
        \EndIf
        \State Update the macro-stress $\bm{\Sigma} = \phi_{\soft}\tstress_{\soft} + \phi_{\hard}\tstress_{\hard}$.
        \State Compute the macroscopic tangent operator $\bm{C}_{\Sigma}$ via Eq.~\eqref{eq:macro-cto}.
    \end{algorithmic}
    \label{algo:material-update}
\end{algorithm}

\subsection{Integration of the viscoplastic model of the soft layer}
As described above, the solution of the bi-layer problem requires the micro-stress tensor and the consistent tangent operator of the individual layers.
For the hard layer, which is linear elastic, these two quantities can be obtained analytically and trivially.
For the soft layer, however, they need to be computed by an integration algorithm for a viscoplastic model of Duvaut--Lions type.

Time integration of a Duvaut--Lions viscoplastic model proceeds in two steps:
(i) integration of the rate-independent plasticity model for the backbone stress tensor,
and (ii) integration of the viscoplastic rate equations for the actual stress tensor and the internal variable(s).
For brevity, we skip an algorithm for the first step---integration of the rate-independent Modified Cam-Clay model---because it is standard and well described in many references (\eg~\cite{Borja2013}).

For the second step, we use the implicit Euler method, except that we evaluate $\tau$ in Eq.~\eqref{eq:relaxation-time} explicitly using $\strain_{\rm v}^{\vpl}$ at the previous time step.
This semi-implicit approach greatly simplifies the integration, without significant compromise in accuracy because $\strain_{\rm v}^{\vpl}$ does not increase dramatically in a load step.
Let us denote the time increment by $\rDelta{t}:=t_{n+1}-t_{n}$.
The viscoplastic rate equations~\eqref{eq:dv-stress-rate} and~\eqref{eq:dv-hardening-rate} are then integrated as
\begin{align}
    \tstrain^{\vpl}_{\soft} &= (\tstrain^{\vpl}_{\soft})_{n}  + \frac{\rDelta{t}}{\tau_{n}}(\mathbb{C}^{\el}_{\soft})^{-1}:(\tstress_{\soft} - \bar{\tstress}_{\soft})\,, \label{eq:dv-stress-rate-discrete}\\
    p_{c} &= (p_{c})_{n} - \frac{\rDelta{t}}{\tau_{n}}(p_{c} - \bar{p}_{c})\,. \label{eq:dv-hardening-rate-discrete}
\end{align}
Rearranging these two equations, we get
\begin{align}
    \tstress_{\soft} &= \frac{\tstress_{\soft}^{\trial} + (\rDelta{t}/\tau_{n})\,\bar{\tstress}_{\soft}}{1 + \rDelta{t}/\tau_{n}}\,, \label{eq:dv-stress-final} \\
    p_{c} &= \frac{(p_{c})_{n} + (\rDelta{t}/\tau_{n})\,\bar{p}_{c}}{1 + \rDelta{t}/\tau_{n}}\,. \label{eq:dv-pc-final}
\end{align}
Here, $\tstress_{\soft}^{\trial} := \mathbb{C}^{\el}_{\soft}:[\tstrain_{\soft} - (\tstrain_{\soft}^{\vpl})_{n}]$ is the trial stress calculated assuming that the strain increment in the load step is fully elastic.
Also, the consistent tangent operator for $\tstress_{\soft}$ can be written as~\cite{Ju1990}
\begin{align}
    \bm{\mathcal{C}}_{\soft} = \frac{\mathbb{C}_{\soft}^{\el} + (\rDelta{t}/\tau_{n})\,\bar{\bm{\mathcal{C}}}_{\soft}}{1 + \rDelta{t}/\tau_{n}}\,,
    \label{eq:dv-cto-final}
\end{align}
where $\bar{\bm{\mathcal{C}}}_{\soft}$ is the consistent tangent operator for the backbone stress $\bar{\tstress}_{\soft}$, which can be obtained by an existing algorithm for the rate-independent model.
The complete procedure for updating the viscoplastic model is summarized in Algorithm~\ref{algo:visco-update}.

\begin{algorithm}[htbp]
    \setstretch{1.15}
    \caption{Viscoplastic update of the soft layer.}
    \begin{algorithmic}[1]
        \Require $\tstrain_{\soft}$ and $\rDelta{t}$
        \Ensure $\tstress_{\soft}$ and $\bm{\mathcal{C}}_{\soft}$

        \State Compute the trial stress $\tstress_{\soft}^{\trial} := \mathbb{C}^{\el}_{\soft}:[\tstrain_{\soft} - (\tstrain_{\soft}^{\vpl})_{n}]$ and set $p_{c} = (p_{c})_{n}$.
        \State Evaluate the yield function $f$ in Eq.~\eqref{eq:yield} with the trial stress.
        \If {$f<0$}
            \State Elastic step. Set $\tstress_{\soft} = \tstress_{\soft}^{\trial}$ and $\bm{\mathcal{C}}_{\soft}=\mathbb{C}_{\soft}^{\el}$.
        \Else
            \State Viscoplastic step. Compute $\bar{\tstress}$ and $\bar{p}_{c}$ using an algorithm for the rate-independent Modified Cam-Clay model.
            \State Calculate $\tstress_{\soft}$ from Eq.~\eqref{eq:dv-stress-final}, $p_{c}$ from Eq.~\eqref{eq:dv-pc-final}, and $\bm{\mathcal{C}}_{\soft}$ from Eq.~\eqref{eq:dv-cto-final}.
        \EndIf
    \end{algorithmic}
    \label{algo:visco-update}
\end{algorithm}

\section{Validation}
\label{sec:validation}
This section validates the proposed model using creep test data on samples from three gas shale formations, namely, Haynesville, Eagle Ford, and Barnett, which have been obtained by Sone and Zoback~\cite{Sone2013b,Sone2014a}.
Each loading stage of the experiments was conducted as follows.
First, a cylindrical sample was subjected to an isotropic confining pressure, which was maintained for $\sim$3 hours.
Subsequently, the sample was axially loaded by a differential stress, which was applied incrementally for 60 seconds and then held constant for $\sim$3 hours.
During the axial loading stage, the sample first experienced instantaneous deformation in the first 60 seconds and then creep deformation for the remaining period.
The values of the confining and differential stresses were varied by test, with complete information available in Sone~\cite{sone2012mechanical}.
In what follows, we simulate these tests using the proposed model and algorithms described above.

\subsection{Haynesville shale}
To begin, we consider the experimental data of the Haynesville-1V sample in Sone and Zoback~\cite{Sone2014a}, which provides the total (instantaneous plus creep) strain in both the axial and lateral directions after axial loading.
The sample was cored along the vertical direction, so it was axially loaded in the direction normal to the bedding plane (\ie~$\theta=0^{\circ}$).
The confining pressure was 30 MPa, and the differential stress was increased by 29 MPa.

Table~\ref{tab:haynesville-parameters} summarizes the model parameters calibrated to reproduce the experimental data of the Haynesville shale sample. These parameters were obtained through the following procedure.
First, the fraction of the soft layer, $\phi_{\soft}$, was calculated as
the average of the clay and kerogen volume fractions of the Haynesville shale samples in Sone and Zoback~\cite{Sone2013a}.
The fraction of the hard layer was then given by $\phi_{\hard} = 1 - \phi_{\soft}$.
The CSL slope, $M$, was computed using the sliding friction coefficient reported in Sone and Zoback~\cite{Sone2013b}
and its relationship with $M$ under triaxial compression.
For the elasticity parameters of the two layers, we first estimated their potential ranges based on the data in Sone and Zoback~\cite{Sone2013a,Sone2013b}, and then further calibrated them to fit the instantaneous strains in the axial and lateral directions.
Then, we calibrated the rate-independent plasticity parameters of the soft layer, initial $p_c$ and $\lambda$, to match the final strain of the experimental data, inferring their possible ranges from the estimated in-situ vertical stress in Sone and Zoback~\cite{Sone2013a} and Modified Cam-Clay parameters for other shales (\eg~\cite{Semnani2016,zhao2018strength,Borja2020}).
It is noted that $p_c$ was also calibrated due to significant uncertainties in the in-situ horizontal stresses and possible sample disturbance, see Haghighat \etal~\cite{Haghighat2020} for a similar treatment.
Lastly, the relaxation time parameters, $\tau_{0}$ and $\zeta$, of the soft layer were calibrated to fit the time evolution of creep strain in the experiment.

\begin{table}[htbp]
    \centering
    \begin{tabular}{lllrr}
    \toprule
    Parameter & Symbol & Units & Soft Layer & Hard Layer \\
    &&& (Viscoplastic MCC) & (Linear elasticity) \\
    \midrule
    Volume fraction & $\phi_{m}$ & -- & 0.455 & 0.545  \\
    Bulk modulus & $K$ & GPa & 5.9 & 41.0 \\
    Poisson's ratio & $\nu$ & -- & 0.28 & 0.36 \\
    CSL slope & $M$ & -- & 0.9 & -- \\
    Preconsolidation pressure & $p_c$ & MPa & 18 & -- \\
    Hardening modulus & $\lambda$ & MPa & 0.0008 & -- \\
    Initial relaxation time & $\tau_0$ & s & 20 & -- \\
    Relaxation time coefficient & $\zeta$ & -- & 3400 & -- \\
    \bottomrule
    \end{tabular}
    \caption{Parameters for Haynesville shale calibrated using experimental data from Sone and Zoback~\cite{Sone2014a}.}
    \label{tab:haynesville-parameters}
\end{table}

\begin{figure}[htbp]
    \centering
    \includegraphics[width=0.6\textwidth]{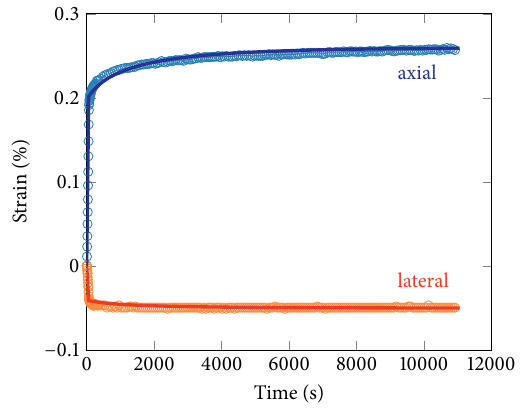}
    \caption{Haynesville shale: experimental data from Sone and Zoback~\cite{Sone2013b} (open circles) and simulation results using the proposed model (solid lines).}
    \label{fig:haynesville}
\end{figure}

Figure~\ref{fig:haynesville} compares the simulation results and the experimental data in terms of the axial and lateral strains from the beginning of axial loading.
The simulation results show an excellent agreement with the experimental data from the instantaneous deformation phase (the first 60 seconds) to the creep phase (the remaining period), in both axial and lateral directions.
It should be emphasized that although an isotropic viscoplasticity model may also be able to mimic the axial strain data, it is unable to accurately simulate the lateral strain data.

\subsection{Eagle Ford shale}
To further validate the proposed model, we use the experimental data of the Eagle Ford-1 samples in Sone and Zoback~\cite{Sone2014a}.
Unlike the previous example, the Eagle Ford data were obtained from two different samples, one cored vertically and the other horizontally.
Thus, the vertical sample was axially loaded in the bedding plane normal direction ($\theta=0^{\circ}$), whereas the horizontal sample was loaded in the bedding plane parallel direction ($\theta=90^{\circ}$).
The confining pressure applied to the two samples were the same (10 MPa, see Sone~\cite{sone2012mechanical}), but the differential stresses were slightly different: 16 MPa for the vertical sample and 17 MPa for the horizontal sample.
The model parameters for the Eagle Ford shale data were obtained through the same procedure as that for the Haynesville shale example and are presented in Table~\ref{tab:eagleford-parameters}.
\begin{table}[htbp]
    \centering
    \begin{tabular}{lllrr}
    \toprule
    Parameter & Symbol & Units & Soft Layer & Hard Layer \\
    &&& (Viscoplastic MCC) & (Linear elasticity) \\
    \midrule
    Volume fraction & $\phi_{m}$ & -- & 0.260 & 0.740  \\
    Bulk modulus & $K$ & GPa & 5.1 & 49.0 \\
    Poisson's ratio & $\nu$ & -- & 0.39 & 0.29 \\
    CSL slope & $M$ & -- & 0.87 & -- \\
    Preconsolidation pressure & $p_c$ & MPa & 18 & -- \\
    Hardening modulus & $\lambda$ & MPa & 0.01 & -- \\
    Initial relaxation time & $\tau_0$ & s & 30 & -- \\
    Relaxation time coefficient & $\zeta$ & -- & 5000 & -- \\
    \bottomrule
    \end{tabular}
    \caption{Parameters for Eagle Ford shale calibrated using experimental data from Sone and Zoback~\cite{Sone2014a}.}
    \label{tab:eagleford-parameters}
\end{table}

\begin{figure}[htbp]
    \centering
    \includegraphics[width=0.55\textwidth]{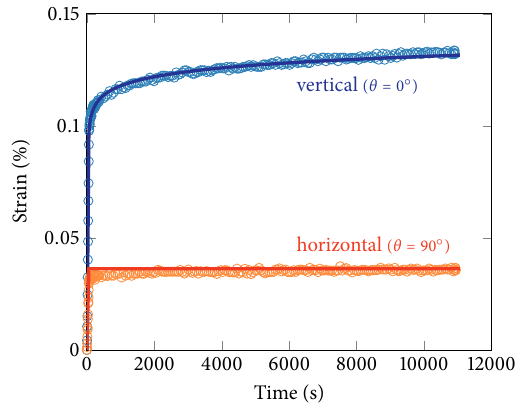}
    \caption{Eagle Ford shale: experimental data from Sone and Zoback~\cite{Sone2014a} (open circles) and simulation results using the proposed model (solid lines). Note that the differential stresses for the vertical and horizontal samples are different: 16 MPa and 17 MPa, respectively.}
    \label{fig:eagleford}
\end{figure}

In Fig.~\ref{fig:eagleford}, the simulation results are compared with the experimental data of the vertical and horizontal Eagle Ford shale samples.
The comparison shows that the proposed model captures the different deformation responses of the vertical and horizontal samples very well.
Remarkably, it can be seen the bedding plane exerts dominant controls over not only the amount of instantaneous deformation but also that of creep.
To incorporate these controls into numerical analysis of shale, one must use an anisotropic viscoplasticity model such as the proposed model.

\subsection{Barnett shale}
As our last validation example, we use the creep test results of the Barnett-1H shale sample in Sone and Zoback~\cite{Sone2013b}.
The sample was cored horizontally, and it was subjected to an isotropic stress of 20 MPa followed by an increase in differential stress of 48 MPa.
Because Sone and Zoback~\cite{Sone2013b} reported creep strains only (\ie~without instantaneous strains), the model parameters are calibrated to the creep strains in the axial and lateral directions.
\revised{The lateral strain in this case is defined as the average of radial strains in the bedding plane parallel and normal directions, in the same way as in the experimental data of Sone and Zoback~\cite{Sone2013b}.}
The calibrated parameters for the Barnett shale data are presented in Table~\ref{tab:barnett-parameters}.
\begin{table}[htbp]
    \centering
    \begin{tabular}{lllrr}
    \toprule
    Parameter & Symbol & Units & Soft Layer & Hard Layer \\
    &&& (Viscoplastic MCC) & (Linear elasticity) \\
    \midrule
    Volume fraction & $\phi_{m}$ & -- & 0.470 & 0.530  \\
    Bulk modulus & $K$ & GPa & 5.9 & 43.0 \\
    Poisson's ratio & $\nu$ & -- & 0.12 & 0.25 \\
    CSL slope & $M$ & -- & 1.2 & -- \\
    Preconsolidation pressure & $p_c$ & MPa & 25 & -- \\
    Hardening modulus & $\lambda$ & MPa & 0.005 & -- \\
    Initial relaxation time & $\tau_0$ & s & 200 & -- \\
    Relaxation time coefficient & $\zeta$ & -- & 5500 & -- \\
    \bottomrule
    \end{tabular}
    \caption{Parameters for Barnett shale calibrated using experimental data from Sone and Zoback~\cite{Sone2013b}.}
    \label{tab:barnett-parameters}
\end{table}

\begin{figure}[htbp]
    \centering
    \includegraphics[width=0.6\textwidth]{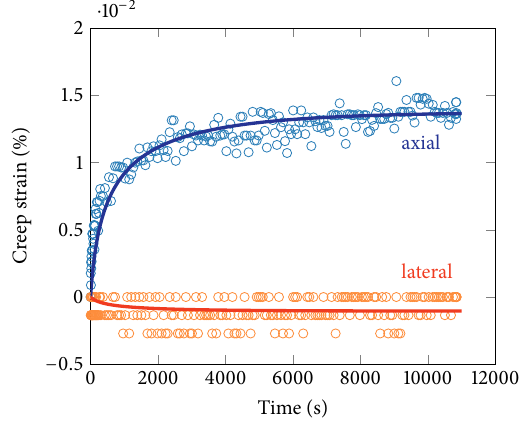}
    \caption{Barnett shale: experimental data from Sone and Zoback~\cite{Sone2013b} (open circles) and simulation results using the proposed model (solid lines).}
    \label{fig:barnett}
\end{figure}

Comparison of the simulation and experimental results in Fig.~\ref{fig:barnett} confirms that the proposed model well reproduces both the axial and lateral creep strains.
For this particular example, it should be noted that the bi-material model of Borja \etal~\cite{Borja2020} has simulated the same experimental data very well.
However, the bi-material model required seven more parameters than the present bi-layer model, mainly because the anisotropic elasto-plastic model used in the bi-material model requires six parameters to represent anisotropy.
Conversely, the present bi-layer model does not require any parameters for anisotropy other than the phase volume fractions, because it accommodates anisotropy through the layered homogenization procedure.  The current procedure may be more computationally expensive, however, due to the need for nested iterations.

\section{Borehole Study}
\label{sec:example}

This section briefly demonstrates the performance of the proposed model for general initial--boundary value problems.
For this purpose, we simulate the the time-dependent closure of a borehole in shale, which is relevant to wellbore stability~\cite{Horsrud1994} and leakage prevention~\cite{Cerasi2017,Xie2019}.
We consider a borehole with a radius of 0.1 m, represented in 2D as a circle inside a square domain in plane strain conditions.
The  shale is modeled using the parameters calibrated for the Haynesville shale sample in the previous section.
To focus on the effect of material anisotropy, we introduce a few simplifying assumptions: the shale is normally consolidated under an isotropic stress field, the borehole is drilled instantaneously, and the shale deforms under fully drained conditions so that no fluid response must be modeled.
All external boundaries are fixed, while the borehole surface is a traction-free boundary.
Using the standard displacement-based finite element method in conjunction with Newton's method,
we solve this problem with 1,944 quadrilateral bilinear elements and a constant time increment of 0.1 minute.
The \texttt{deal.II} library~\cite{BangerthHartmannKanschat2007,dealII91} is used for the finite element solution.
To examine the effect of shale anisotropy on the borehole closure behavior, we repeat the same problem with three different bedding plane directions: $\theta=0^{\circ}$, $\theta=45^{\circ}$, and $\theta=90^{\circ}$.

\begin{figure}[htbp]
  \centering
  \includegraphics[width=0.9\textwidth]{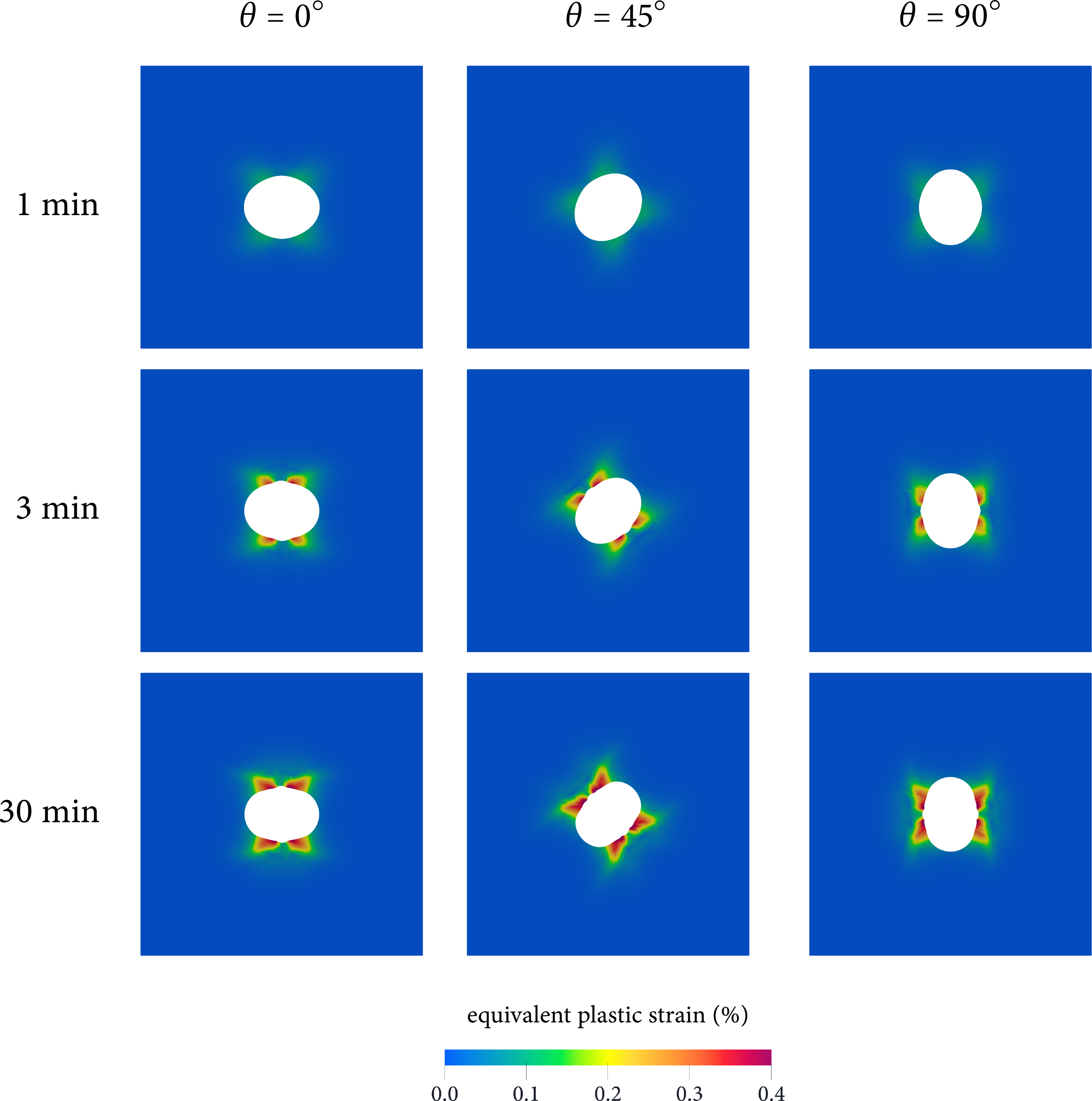}
  \caption{Borehole closure simulation: time evolution of the borehole geometry and equivalent plastic strain field with different bedding plane directions. Displacements are exaggerated by a factor of 200.}
  \label{fig:borehole-eps}
\end{figure}

Figure~\ref{fig:borehole-eps} presents the simulated time evolution of the borehole geometry and the equivalent plastic strain field ($\sqrt{2/3}\,\|\tstrain^{\vpl}\|$).
It can be seen that in all cases the borehole progressively closes over time, but the closure patterns differ according to the bedding plane direction.
The boreholes squeeze more in the bedding plane normal direction than the parallel directions to become oval-shaped, even though the initial geometry and stress condition had no significant directional dependence.
Consequently, the bedding plane direction controls where viscoplastic zones emerge.
The closure process in this problem nears steady-state after 30 minutes, but the real process in the subsurface would be delayed significantly due to the slow dissipation of pore pressure in low-permeability shale.
The fundamental deformation patterns would, however, remain similar.

\begin{figure}[htbp]
  \centering
  \subfloat[$\theta=0^{\circ}$]{\includegraphics[width=0.32\textwidth]{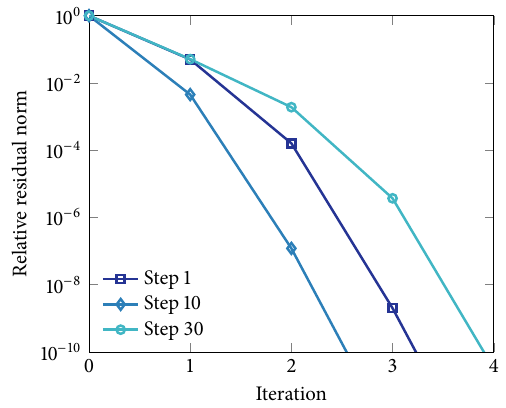}}\hspace{0.5em}
  \subfloat[$\theta=45^{\circ}$]{\includegraphics[width=0.32\textwidth]{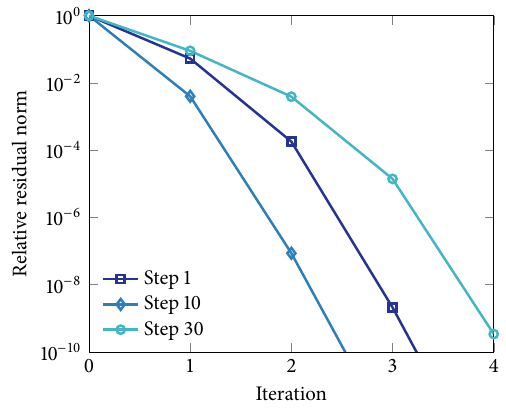}}\hspace{0.5em}
  \subfloat[$\theta=90^{\circ}$]{\includegraphics[width=0.32\textwidth]{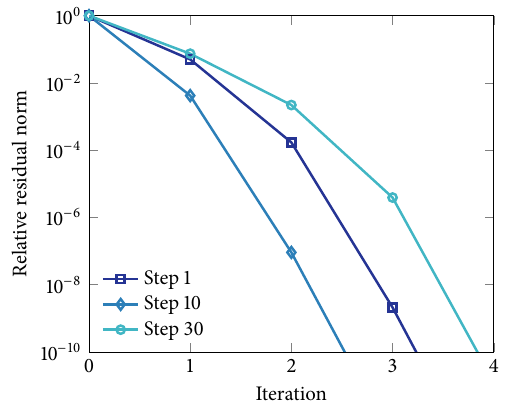}}
  \caption{Borehole closure simulation: convergence profiles of Newton iterations.}
  \label{fig:borehole-convergence}
\end{figure}

To demonstrate the computational efficiency and robustness of the proposed model,
Fig.~\ref{fig:borehole-convergence} presents the convergence profiles during the Newton iterations in a few earlier steps where viscoplastic deformations rapidly develop.
As can be seen, the relative residual norms are quickly reduced to small values after only four Newton iterations.
While not presented, as the viscoplastic process nears completion, the Newton iterations converge even faster.
The convergence behavior is near-optimal in every step, confirming the correctness of the macroscopic and microscopic tangent operators.

The results of this example highlight that material anisotropy alone has a profound impact on borehole deformations, and that the proposed model can be easily incorporated into a standard numerical code for initial--boundary value problems.
It is also noted that the model could be incorporated into more sophisticated numerical methods for coupled poromechanical problems (\eg~\cite{choo2015stabilized,choo2019stabilized,zhang2019preferential,camargo2020macroelement}) to account for fluid flow effects on viscoplastic deformation.
Future work will apply the model to investigate the combined control of the bedding plane direction, stress anisotropy, and fluid flow on the time-dependent behavior of boreholes in shale.

\section{Closure}
\label{sec:closure}

This work has proposed an anisotropic viscoplasticity model for shale, extending a recently developed homogenization framework for layered materials~\cite{Semnani2020}.
The proposed model conceptualizes the microstructure of shale as a composite of a viscoplastic layer and an elastic layer.
The homogenization approach allows the individual layers to be described by familiar isotropic constitutive models, with anisotropy emerging as an inherent property of the layered microstructure.
This combination of layer-wise isotropic models also helps minimize the number of free material parameters, all of which have clear physical meanings and can be determined without significant difficulty.
The proposed model was validated using experimental creep data from three gas shale formations, showing good agreement for all strain components.
We have also demonstrated that the model can be efficiently incorporated within standard numerical workflows, with simulation of a borehole closure process as a representative example.
The remarkable success of the proposed model suggests that homogenization is a powerful approach to the modeling of physical behavior in layered geomaterials, which remains a difficult challenge to accurately predicting the response of geomechanical systems.

\section*{Acknowledgments}
JC acknowledges financial support from the Research Grants Council of Hong Kong under Project 27205918.
SJS and JAW were supported by Total S.A. through the FC-MAELSTROM project. Portions of this work were performed under the auspices of the U.S. Department of Energy by Lawrence Livermore National Laboratory under Contract DE-AC52-07-NA27344.
The authors wish to thank Yashar Mehmani and Alan Burnham for providing Figure 1 and for helpful discussions regarding shale anisotropy.
\revised{The authors are also grateful to Hiroki Sone for sharing his experimental data and for valuable discussions regarding the experimental procedure.}

\bibliography{references}

\begin{thebibliography}{10}
\expandafter\ifx\csname url\endcsname\relax
  \def\url#1{\texttt{#1}}\fi
\expandafter\ifx\csname urlprefix\endcsname\relax\def\urlprefix{URL }\fi
\expandafter\ifx\csname href\endcsname\relax
  \def\href#1#2{#2} \def\path#1{#1}\fi

\bibitem{Sone2014a}
H.~Sone, M.~D. Zoback, Time-dependent deformation of shale gas reservoir rocks
  and its long-term effect on the in situ state of stress, International
  Journal of Rock Mechanics and Mining Sciences 69 (2014) 120--132.

\bibitem{Sone2014b}
H.~Sone, M.~D. Zoback, Viscous relaxation model for predicting least principal
  stress magnitudes in sedimentary rocks, Journal of Petroleum Science and
  Engineering 124 (2014) 416--431.

\bibitem{Horsrud1994}
P.~Horsrud, R.~M. Holt, E.~F. Sonstebo, G.~Svano, B.~Bostrom, Time dependent
  borehole stability: laboratory studies and numerical simulation of different
  mechanisms in shale, in: Rock Mechanics in Petroleum Engineering, Society of
  Petroleum Engineers, 1994, pp. SPE--28060--MS.

\bibitem{Cerasi2017}
P.~Cerasi, E.~Lund, M.~L. Kleiven, A.~Stroisz, S.~Pradhan, C.~Kj{\o}ller,
  P.~Frykman, E.~Fj{\ae}r, {Shale creep as leakage healing mechanism in CO$_2$
  sequestration}, Energy Procedia 114 (2017) 3096--3112.

\bibitem{Xie2019}
X.~Xie, E.~Fj{\ae}r, E.~Detournay, Time-dependent closure of a borehole in a
  viscoplastic rock, Geomechanics for Energy and the Environment 19 (2019)
  100115.

\bibitem{Asala2016}
H.~Asala, M.~Ahmadi, A.~Taleghani, Why re-fracturing works and under what
  conditions, in: SPE Annual Technical Conference and Exhibition, Society of
  Petroleum Engineers, 2016, pp. SPE--181516--MS.

\bibitem{Yang2016}
Y.~Yang, M.~Zoback, {Viscoplastic deformation of the Bakken and adjacent
  formations and its relation to hydraulic fracture growth}, Rock Mechanics and
  Rock Engineering 49 (2016) 689--698.

\bibitem{Chau2017}
V.~Chau, C.~Li, S.~Rahimi-Aghdam, Z.~Ba\v{z}ant, {The enigma of large-scale
  permeability of gas shale: Pre-existing or frac-induced?}, Journal of Applied
  Mechanics 84~(6) (2017).

\bibitem{Chang2009}
C.~Chang, M.~D. Zoback, {Viscous creep in room-dried unconsolidated Gulf of
  Mexico shale (I): Experimental results}, Journal of Petroleum Science and
  Engineering 69~(3--4) (2009) 239--246.

\bibitem{Li2012CreepShale}
Y.~Li, A.~Ghassemi, {Creep behavior of Barnett, Haynesville, and Marcellus
  shale}, in: 46th US Rock Mechanics/Geomechanics Symposium, American Rock
  Mechanics Association, 2012.

\bibitem{Sone2013b}
H.~Sone, M.~D. Zoback, {Mechanical properties of shale-gas reservoir rocks --
  Part 2: Ductile creep, brittle strength, and their relation to the elastic
  modulus}, Geophysics 78~(5) (2013) D393--D402.

\bibitem{Bennett2015}
K.~C. Bennett, L.~A. Berla, W.~D. Nix, R.~I. Borja, {Instrumented
  nanoindentation and 3D mechanistic modeling of a shale at multiple scales},
  Acta Geotechnica 10~(1) (2015) 1--14.

\bibitem{Rybacki2017}
E.~Rybacki, J.~Herrmann, R.~Wirth, G.~Dresen, {Creep of Posidonia shale at
  elevated pressure and temperature}, Rock Mechanics and Rock Engineering 50
  (2017) 3121--3140.

\bibitem{geng2018time}
Z.~Geng, A.~Bonnelye, M.~Chen, Y.~Jin, P.~Dick, C.~David, X.~Fang, A.~Schubnel,
  {Time and temperature dependent creep in Tournemire shale}, Journal of
  Geophysical Research: Solid Earth 123~(11) (2018) 9658--9675.

\bibitem{rassouli2018comparison}
F.~S. Rassouli, M.~D. Zoback, Comparison of short-term and long-term creep
  experiments in shales and carbonates from unconventional gas reservoirs, Rock
  Mechanics and Rock Engineering 51~(7) (2018) 1995--2014.

\bibitem{Trzeciak2018}
M.~Trzeciak, H.~Sone, M.~Dabrowski, {Long-term creep tests and viscoelastic
  constitutive modeling of lower Paleozoic shales from the Baltic Basin, N
  Poland}, International Journal of Rock Mechanics and Mining Sciences 112
  (2018) 139--157.

\bibitem{Slim2019}
M.~Slim, S.~Abedi, L.~T. Bryndzia, F.-J. Ulm, Role of organic matter on
  nanoscale and microscale creep properties of source rocks, Journal of
  Engineering Mechanics 145~(1) (2019) 04018121.

\bibitem{Mighani2019CreepExperiments}
S.~Mighani, Y.~Bernab{\'{e}}, A.~Boulenouar, U.~Mok, B.~Evans, {Creep
  deformation in Vaca Muerta shale from nanoindentation to triaxial
  experiments}, Journal of Geophysical Research: Solid Earth 124~(8) (2019)
  7842--7868.

\bibitem{Herrmann2020}
J.~Herrmann, E.~Rybacki, H.~Sone, G.~Dresen, {Deformation experiments on
  Bowland and Posidonia shale---Part II: Creep behavior at in istu $p_c$--$T$
  conditions}, Rock Mechanics and Rock Engineering 53 (2019) 755--779.

\bibitem{Li2020}
C.~Li, J.~Wang, H.~Xie, Anisotropic creep characteristics and mechanism of
  shale under elevated deviatoric stress, Journal of Petroleum Science and
  Engineering 185 (2020) 106670.

\bibitem{Mehmani2016}
Y.~Mehmani, A.~K. Burnham, M.~D.~V. Berg, F.~Gelin, H.~Tchelepi, Quantification
  of kerogen content in organic-rich shales from optical photographs, Fuel 177
  (2016) 63--75.

\bibitem{Trzeciak2018Long-termPoland}
M.~Trzeciak, H.~Sone, M.~Dabrowski, {Long-term creep tests and viscoelastic
  constitutive modeling of lower Paleozoic shales from the Baltic Basin, N
  Poland}, International Journal of Rock Mechanics and Mining Sciences 112
  (2018) 139--157.

\bibitem{DUSSEAULT1993Time-dependentRocks}
M.~B. Dusseault, C.~J. Fordham, {Time-dependent behavior of rocks}, in: Rock
  Testing and Site Characterization, Elsevier, 1993, pp. 119--149.

\bibitem{Zhou2005ElasticTests}
C.~Zhou, J.-H. Yin, J.-G. Zhu, C.-M. Cheng, {Elastic anisotropic viscoplastic
  modeling of the strain-rate-dependent stress–strain behavior of
  $K_0$-consolidated natural marine clays in triaxial shear tests},
  International Journal of Geomechanics 5~(3) (2005) 218--232.

\bibitem{Sivasithamparam2015ModellingSoils}
N.~Sivasithamparam, M.~Karstunen, P.~Bonnier, {Modelling creep behaviour of
  anisotropic soft soils}, Computers and Geotechnics 69 (2015) 46--57.

\bibitem{Jiang2017EvaluationClays}
J.~Jiang, H.~I. Ling, V.~N. Kaliakin, X.~Zeng, C.~Hung, {Evaluation of an
  anisotropic elastoplastic--viscoplastic bounding surface model for clays},
  Acta Geotechnica 12~(2) (2017) 335--348.

\bibitem{Leoni2008AnisotropicSoils}
M.~Leoni, M.~Karstunen, P.~A. Vermeer, {Anisotropic creep model for soft
  soils}, G\'{e}otechnique 58~(3) (2008) 215--226.

\bibitem{Yin2010AnClays}
Z.~Y. Yin, C.~S. Chang, M.~Karstunen, P.~Y. Hicher, {An anisotropic
  elastic-viscoplastic model for soft clays}, International Journal of Solids
  and Structures 47~(5) (2010) 665--677.

\bibitem{Karim2014ReviewClays}
M.~R. Karim, C.~T. Gnanendran, {Review of constitutive models for describing
  the time dependent behaviour of soft clays}, Geomechanics and Geoengineering
  9~(1) (2014) 36--51.

\bibitem{roscoe1968}
K.~Roscoe, J.~Burland, On the generalized stress-strain behaviour of `wet'
  clay, in: J.~Heyman, F.~Leckie (Eds.), Engineering Plasticity, Cambridge
  University Press, Cambridge, 1968, pp. 535--609.

\bibitem{Pietruszczak2004DescriptionMaterials}
S.~Pietruszczak, D.~Lydzba, J.-F. Shao, {Description of creep in inherently
  anisotropic frictional materials}, Journal of Engineering Mechanics 130~(6)
  (2004) 681--690.

\bibitem{Borja2020}
R.~I. Borja, Q.~Yin, Y.~Zhao, {Cam-Clay plasticity. Part IX: On the anisotropy,
  heterogeneity, and viscoplasticity of shale}, Computer Methods in Applied
  Mechanics and Engineering 360 (2020) 112695.

\bibitem{Semnani2016}
S.~J. Semnani, J.~A. White, R.~I. Borja, Thermoplasticity and strain
  localization in transversely isotropic materials based on anisotropic
  critical state plasticity, International Journal for Numerical and Analytical
  Methods in Geomechanics 40 (2016) 2423--2449.

\bibitem{Yu2002MultiscaleProblem}
Q.~Yu, J.~Fish, {Multiscale asymptotic homogenization for multiphysics problems
  with multiple spatial and temporal scales: a coupled thermo-viscoelastic
  example problem}, International journal of solids and structures 39~(26)
  (2002) 6429--6452.

\bibitem{Lahellec2007EffectiveApproach}
N.~Lahellec, P.~Suquet, {Effective behavior of linear viscoelastic composites:
  A time-integration approach}, International Journal of Solids and Structures
  44~(2) (2007) 507--529.

\bibitem{Fotiu1996OverallComposites}
P.~A. Fotiu, S.~Nemat-Nasser, {Overall properties of elastic-viscoplastic
  periodic composites}, International Journal of Plasticity 12~(2) (1996)
  163--190.

\bibitem{Nesenenko2007HomogenizationViscoplasticity}
S.~Nesenenko, {Homogenization in viscoplasticity}, SIAM Journal on Mathematical
  Analysis 39~(1) (2007) 236--262.

\bibitem{Nesenenko2013HomogenizationType}
S.~Nesenenko, {Homogenization of rate-dependent inelastic models of monotone
  type}, Asymptotic Analysis 81~(1) (2013) 1--29.

\bibitem{Mercier2009HomogenizationSchemes}
S.~Mercier, A.~Molinari, {Homogenization of elastic-viscoplastic heterogeneous
  materials: Self-consistent and Mori--Tanaka schemes}, International Journal
  of Plasticity 25~(6) (2009) 1024--1048.

\bibitem{Suquet1987ElementsMechanics}
P.~M. Suquet, {Elements of Homogenization for Inelastic Solid Mechanics}, in:
  E.~Sanchez-Palencia, A.~Zaoui (Eds.), Homogenization techniques for composite
  media, Springer-Verlag, 1987, pp. 193--278.

\bibitem{Charalambakis2018MathematicalProgress}
N.~Charalambakis, G.~Chatzigeorgiou, Y.~Chemisky, F.~Meraghni, {Mathematical
  homogenization of inelastic dissipative materials: a survey and recent
  progress}, Continuum Mechanics and Thermodynamics 30~(1) (2018) 1--51.

\bibitem{Berbenni2015AExtension}
S.~Berbenni, L.~Capolungo, {A Mori-Tanaka homogenization scheme for non-linear
  elasto-viscoplastic heterogeneous materials based on translated fields: An
  affine extension}, Comptes Rendus M{\'{e}}canique 343~(2) (2015) 95--106.

\bibitem{Paquin1999IntegralMaterials}
A.~Paquin, H.~Sabar, M.~Berveiller, {Integral formulation and self-consistent
  modelling of elastoviscoplastic behavior of heterogeneous materials}, Archive
  of Applied Mechanics 69~(1) (1999) 14--35.

\bibitem{Dvorak1992TransformationMaterials}
G.~J. Dvorak, {Transformation field analysis of inelastic composite materials},
  Proc. R. Soc. Lond. A 437 (1992) 311--327.

\bibitem{Eshelby1961}
J.~D. Eshelby, Elastic inclusions and inhomogeneities, in: I.~Sneddon, R.~Hill
  (Eds.), Progress in Solid Mechanics, North-Holland, Amsterdam, 1961, pp.
  89--140.

\bibitem{Abou-ChakraGuery2008AGeomaterial}
A.~A.-C. Gu{\'e}ry, F.~Cormery, K.~Su, J.-F. Shao, D.~Kondo, A micromechanical
  model for the elasto-viscoplastic and damage behavior of a cohesive
  geomaterial, Physics and Chemistry of the Earth, Parts A/B/C 33 (2008)
  S416--S421.

\bibitem{Pierard2006AnComposites}
O.~Pierard, I.~Doghri, {An enhanced affine formulation and the corresponding
  numerical algorithms for the mean-field homogenization of elasto-viscoplastic
  composites}, International Journal of Plasticity 22~(1) (2006) 131--157.

\bibitem{Doghri2010Mean-fieldMethod}
I.~Doghri, L.~Adam, N.~Bilger, {Mean-field homogenization of
  elasto-viscoplastic composites based on a general incrementally affine
  linearization method}, International Journal of Plasticity 26~(2) (2010)
  219--238.

\bibitem{Lahellec2007OnPrinciples}
N.~Lahellec, P.~Suquet, {On the effective behavior of nonlinear inelastic
  composites: I. Incremental variational principles}, Journal of the Mechanics
  and Physics of Solids 55~(9) (2007) 1932--1963.

\bibitem{Brassart2012HomogenizationPrinciple}
L.~Brassart, L.~Stainier, I.~Doghri, L.~Delannay, {Homogenization of
  elasto-(visco) plastic composites based on an incremental variational
  principle}, International Journal of Plasticity 36 (2012) 86--112.

\bibitem{Boudet2016AnComposites}
J.~Boudet, F.~Auslender, M.~Bornert, Y.~Lapusta, {An incremental variational
  formulation for the prediction of the effective work-hardening behavior and
  field statistics of elasto-(visco)plastic composites}, International Journal
  of Solids and Structures 83 (2016) 90--113.

\bibitem{Brassart2011HomogenizationApproaches}
L.~Brassart, {Homogenization of elasto-(visco)plastic composites:
  history-dependent incremental and variational approaches}, Ph.D. thesis,
  {Universit\'{e} catholique de Louvaino} (2011).

\bibitem{Molinari2002AveragingMaterials}
A.~Molinari, {Averaging models for heterogeneous viscoplastic and elastic
  viscoplastic materials}, Journal of Engineering Materials and Technology,
  Transactions of the ASME 124~(1) (2002) 62--70.

\bibitem{Mercier2009HomogenizationSchemesb}
S.~Mercier, A.~Molinari, {Homogenization of elastic-viscoplastic heterogeneous
  materials: Self-consistent and Mori-Tanaka schemes}, International Journal of
  Plasticity 25~(6) (2009) 1024--1048.

\bibitem{Mareau2009MicromechanicalMaterials}
C.~Mareau, V.~Favier, M.~Berveiller, {Micromechanical modeling coupling
  time-independent and time-dependent behaviors for heterogeneous materials},
  International Journal of Solids and Structures 46~(2) (2009) 223--237.

\bibitem{Mareau2015AnMethod}
C.~Mareau, S.~Berbenni, {An affine formulation for the self-consistent modeling
  of elasto-viscoplastic heterogeneous materials based on the translated field
  method}, International Journal of Plasticity 64 (2015) 134--150.

\bibitem{Sabar2002AMaterials}
H.~Sabar, M.~Berveiller, V.~Favier, S.~Berbenni, {A new class of micro--macro
  models for elastic--viscoplastic heterogeneous materials}, International
  Journal of Solids and Structures 39~(12) (2002) 3257--3276.

\bibitem{Marfia2018MultiscaleComposites}
S.~Marfia, E.~Sacco, {Multiscale technique for nonlinear analysis of
  elastoplastic and viscoplastic composites}, Composites Part B: Engineering
  136 (2018) 241--253.

\bibitem{Covezzi2017HomogenizationTFA}
F.~Covezzi, S.~de~Miranda, S.~Marfia, E.~Sacco, {Homogenization of
  elastic--viscoplastic composites by the Mixed TFA}, Computer Methods in
  Applied Mechanics and Engineering 318 (2017) 701--723.

\bibitem{Roussette2009NonuniformComposites}
S.~Roussette, J.~C. Michel, P.~Suquet, {Nonuniform transformation field
  analysis of elastic-viscoplastic composites}, Composites Science and
  Technology 69~(1) (2009) 22--27.

\bibitem{Kruch2011Multi-scaleDamage}
S.~Kruch, J.-L. Chaboche, {Multi-scale analysis in elasto-viscoplasticity
  coupled with damage}, International Journal of Plasticity 27~(12) (2011)
  2026--2039.

\bibitem{Barral2020HomogenizationValidation}
M.~Barral, G.~Chatzigeorgiou, F.~Meraghni, R.~L{\'{e}}on, {Homogenization using
  modified Mori-Tanaka and TFA framework for
  elastoplastic-viscoelastic-viscoplastic composites: Theory and numerical
  validation}, International Journal of Plasticity 127 (2020).

\bibitem{Ohno2000HomogenizedStructures}
N.~Ohno, X.~Wu, T.~Matsuda, {Homogenized properties of elastic-viscoplastic
  composites with periodic internal structures}, International Journal of
  Mechanical Sciences 42~(8) (2000) 1519--1536.

\bibitem{Nakata2008Multi-scaleConfigurationb}
K.~Nakata, T.~Matsuda, M.~Kawai, {Multi-scale creep analysis of plain-woven
  laminates using time-dependent homogenization theory: Effects of laminate
  configuration}, International Journal of Modern Physics B 22~(31n32) (2008)
  6173--6178.

\bibitem{Ohno2001ADistributions}
N.~Ohno, T.~Matsuda, X.~Wu, {A homogenization theory for elastic-viscoplastic
  composites with point symmetry of internal distributions}, International
  Journal of Solids and Structures 3 38 (2001) 2867--2878.

\bibitem{Chung2000AMedia}
P.~W. Chung, K.~K. Tamma, R.~R. Namburu, {A micro/macro homogenization approach
  for viscoelastic creep analysis with dissipative correctors for heterogeneous
  woven-fabric layered media}, Composites science and technology 60~(12-13)
  (2000) 2233--2253.

\bibitem{Fish1998ComputationalHomogenization}
J.~Fish, K.~Shek, {Computational damage mechanics for composite materials based
  on mathematical homogenization}, International Journal for Numerical Methods
  in Engineering 45 (1998) 1657--1679.

\bibitem{Chatzigeorgiou2016PeriodicMaterialsb}
G.~Chatzigeorgiou, N.~Charalambakis, Y.~Chemisky, F.~Meraghni, {Periodic
  homogenization for fully coupled thermomechanical modeling of dissipative
  generalized standard materials}, International Journal of Plasticity 81
  (2016) 18--39.

\bibitem{Matsuda2002HomogenizedLaminates}
T.~Matsuda, N.~Ohno, H.~Tanaka, T.~Shimizu, {Homogenized in-plane
  elastic-viscoplastic behavior of long fiber-reinforced laminates}, JSME
  International Journal 45~(4) (2002) 538--544.

\bibitem{Shamaev2016HomogenizationMaterials}
A.~S. Shamaev, V.~V. Shumilova, {Homogenization of the equations of state for a
  heterogeneous layered medium consisting of two creep materials}, Proceedings
  of the Steklov Institute of Mathematics 295~(1) (2016) 213--224.

\bibitem{Semnani2020}
S.~J. Semnani, J.~A. White, An inelastic homogenization framework for layered
  materials with planes of weakness, Computer Methods in Applied Mechanics and
  Engineering 370 (2020) 113221.

\bibitem{Haghighat2020}
E.~Haghighat, F.~S. Rassouli, M.~D. Zoback, R.~Juanes, A viscoplastic model of
  creep in shale, Geophysics 85~(3) (2020) MR155--MR166.

\bibitem{Xu2017UtilizingBasin}
S.~Xu, F.~S. Rassouli, M.~D. Zoback, {Utilizing a viscoplastic stress
  relaxation model to study vertical hydraulic fracture propagation in Permian
  Basin}, in: SPE/AAPG/SEG Unconventional Resources Technology Conference 2017,
  Unconventional Resources Technology Conference (URTEC), 2017.

\bibitem{navarro2001secondary}
V.~Navarro, E.~Alonso, Secondary compression of clays as a local dehydration
  process, G{\'e}otechnique 51~(10) (2001) 859--869.

\bibitem{cosenza2014secondary}
P.~Cosenza, D.~Koro{\v{s}}ak, Secondary consolidation of clay as an anomalous
  diffusion process, International Journal for Numerical and Analytical Methods
  in Geomechanics 38~(12) (2014) 1231--1246.

\bibitem{choo2016hydromechanical}
J.~Choo, J.~A. White, R.~I. Borja, Hydromechanical modeling of unsaturated flow
  in double porosity media, International Journal of Geomechanics 16~(6) (2016)
  D4016002.

\bibitem{borja2016cam}
R.~I. Borja, J.~Choo, {Cam-Clay plasticity, Part VIII: A constitutive framework
  for porous materials with evolving internal structure}, Computer Methods in
  Applied Mechanics and Engineering 309 (2016) 653--679.

\bibitem{Choo2011}
J.~Choo, Y.-H. Jung, C.-K. Chung, {Effect of directional stress history on
  anisotropy of initial stiffness of cohesive soils measured by bender element
  tests}, Soils and foundations 51~(4) (2011) 737--747.

\bibitem{Choo2013}
J.~Choo, Y.-H. Jung, W.~Cho, C.-K. Chung, {Effect of pre-shear stress path on
  nonlinear shear stiffness degradation of cohesive soils}, Geotechnical
  Testing Journal 36~(2) (2013) 198--205.

\bibitem{Duvaut1972}
G.~Duvaut, J.~Lions, Inequalities in Mechanics and Physics, Springer, 1972.

\bibitem{Simo1988}
J.~C. Simo, J.~G. Kennedy, S.~Govindjee, Non-smooth multisurface plasticity and
  viscoplasticity. loading/unloading conditions and numerical algorithms,
  International Journal for Numerical Methods in Engineering 26~(10) (1988)
  2161--2185.

\bibitem{Ju1990}
J.~W. Ju, Consistent tangent moduli for a class of viscoplasticity, Journal of
  Engineering Mechanics 116~(8) (1990) 1764--1779.

\bibitem{wang1997viscoplasticity}
W.~Wang, L.~Sluys, R.~De~Borst, Viscoplasticity for instabilities due to strain
  softening and strain-rate softening, International Journal for Numerical
  Methods in Engineering 40~(20) (1997) 3839--3864.

\bibitem{lazari2015local}
M.~Lazari, L.~Sanavia, B.~Schrefler, Local and non-local elasto-viscoplasticity
  in strain localization analysis of multiphase geomaterials, international
  Journal for Numerical and Analytical methods in Geomechanics 39~(14) (2015)
  1570--1592.

\bibitem{perzyna1966fundamental}
P.~Perzyna, Fundamental problems in viscoplasticity, in: Advances in applied
  mechanics, Vol.~9, Elsevier, 1966, pp. 243--377.

\bibitem{Chang2010}
C.~Chang, M.~D. Zoback, {Viscous creep in room-dried unconsolidated Gulf of
  Mexico shale (II): Development of a viscoplasticity model}, Journal of
  Petroleum Science and Engineering 72~(1--2) (2010) 50--55.

\bibitem{white2017thermoplasticity}
J.~A. White, A.~K. Burnham, D.~W. Camp, A thermoplasticity model for oil shale,
  Rock Mechanics and Rock Engineering 50~(3) (2017) 677--688.

\bibitem{zhao2018strength}
Y.~Zhao, S.~J. Semnani, Q.~Yin, R.~I. Borja, On the strength of transversely
  isotropic rocks, International Journal for Numerical and Analytical Methods
  in Geomechanics 42~(16) (2018) 1917--1934.

\bibitem{Backus1962}
G.~E. Backus, Long-wave elastic anisotropy produced by horizontal layering,
  Journal of Geophysical Research 67~(11) (1962) 4427--4440.

\bibitem{Borja2013}
R.~I. Borja, {Plasticity Modeling and Computation}, Springer, 2013.

\bibitem{sone2012mechanical}
H.~Sone, Mechanical properties of shale gas reservoir rocks, and its relation
  to the in-situ stress variation observed in shale gas reservoirs, Ph.D.
  thesis, Stanford university (2012).

\bibitem{Sone2013a}
H.~Sone, M.~D. Zoback, {Mechanical properties of shale-gas reservoir rocks --
  Part 1: Static and dynamic elastic properties and anisotropy}, Geophysics
  78~(5) (2013) D381--D392.

\bibitem{BangerthHartmannKanschat2007}
W.~Bangerth, R.~Hartmann, G.~Kanschat, {deal.II} -- a general purpose object
  oriented finite element library, ACM Trans. Math. Softw. 33~(4) (2007)
  24/1--24/27.

\bibitem{dealII91}
D.~Arndt, W.~Bangerth, T.~C. Clevenger, D.~Davydov, M.~Fehling,
  D.~Garcia-Sanchez, G.~Harper, T.~Heister, L.~Heltai, M.~Kronbichler, R.~M.
  Kynch, M.~Maier, J.-P. Pelteret, B.~Turcksin, D.~Wells, The \texttt{deal.II}
  library, version 9.1, Journal of Numerical Mathematics 27~(4) (2019)
  203--213.

\bibitem{choo2015stabilized}
J.~Choo, R.~I. Borja, Stabilized mixed finite elements for deformable porous
  media with double porosity, Computer Methods in Applied Mechanics and
  Engineering 293 (2015) 131--154.

\bibitem{choo2019stabilized}
J.~Choo, {Stabilized mixed continuous/enriched Galerkin formulations for
  locally mass conservative poromechanics}, Computer Methods in Applied
  Mechanics and Engineering 357 (2019) 112568.

\bibitem{zhang2019preferential}
Q.~Zhang, J.~Choo, R.~I. Borja, {On the preferential flow patterns induced by
  transverse isotropy and non-Darcy flow in double porosity media}, Computer
  Methods in Applied Mechanics and Engineering 353 (2019) 570--592.

\bibitem{camargo2020macroelement}
J.~T. Camargo, J.~A. White, R.~I. Borja, A macroelement stabilization for mixed
  finite element/finite volume discretizations of multiphase poromechanics,
  Computational Geosciences (2020) 1--18\href
  {https://doi.org/10.1007/s10596-020-09964-3}
  {\path{doi:10.1007/s10596-020-09964-3}}.

\end{thebibliography}

\end{document}